\documentclass[reprint,aps,prl,nofootinbib]{revtex4-1}
\usepackage{graphicx}
\usepackage[utf8]{inputenc}
\usepackage{amsmath}
\usepackage{natbib}
\usepackage{bm}
\usepackage{siunitx}
\usepackage{placeins}
\usepackage[caption=false]{subfig}
\usepackage{amsmath}
\usepackage{color}
\graphicspath{{./images/}}
\usepackage{textcomp}

\begin{document}
\title{Electrical transport and optical band gap of NiFe$_\textrm{2}$O$_\textrm{x}$ thin films}
\author{Panagiota Bougiatioti$^1$, Orestis Manos$^1$, Christoph Klewe$^{1,2}$, Daniel Meier$^1$, Niclas Teichert$^1$, Jan-Michael Schmalhorst$^1$, Timo~Kuschel$^{1}$, and G\"unter~Reiss$^1$, \email{Electronic mail: pbougiatioti@physik.uni-bielefeld.de}}
\affiliation{$^1$\mbox{Center for Spinelectronic Materials and Devices, Department of Physics,}\\
\mbox{Bielefeld University, Universit\"atsstra\ss e 25, 33615 Bielefeld, Germany}\\
$^2$\mbox{Advanced Light Source, Lawrence Berkeley National Laboratory, California 94720, USA}\\}

\date{\today}


\begin{abstract}

We fabricated NiFe$_\textrm{2}$O$_\textrm{x}$ thin films on MgAl$_2$O$_4$(001) substrates by reactive dc magnetron co-sputtering varying the oxygen partial pressure during deposition. The fabrication of a variable material with oxygen deficiency leads to controllable electrical and optical properties which would be beneficial for the investigations of the transport phenomena and would, therefore, promote the use of such materials in spintronic and spin caloritronic applications. We used several characterization techniques in order to investigate the film properties, focusing on their structural, magnetic, electrical, and optical properties. From the electrical resistivity measurements we obtained the conduction mechanisms that govern the systems in high and low temperature regimes, extracting low thermal activation energies which unveil extrinsic transport mechanisms. The thermal activation energy decreases in the less oxidized samples revealing the pronounced contribution of a large amount of electronic states localized in the band gap to the electrical conductivity. Hall effect measurements showed the mixed-type semiconducting character of our films. The optical band gaps were determined via ultraviolet-visible spectroscopy. They follow a similar trend as the thermal activation energy, with lower band gap values in the less oxidized samples.

\end{abstract}

\maketitle

\section{Introduction}

The recently established field of spin caloritronics \cite{Hoffmann:2015} combines research on spin-related phenomena with thermoelectric effects. After the observation of phenomena such as the spin Hall effect \cite{Hirsch:1999} and the spin Seebeck effect \cite{Uchida:2008,Uchida:2010,Uchida:2014} a plethora of investigations have been reported towards the generation, manipulation and detection of pure spin currents in ferro(i)magnetic (FM) materials. For the investigation of such transport phenomena in conducting or semiconducting materials possible contributions from additional effects should be taken into account. For example, in case of studying the longitudinal spin Seebeck effect (LSSE) in metallic or semiconducting FMs, an additional anomalous Nernst effect (ANE) contribution can contaminate the LSSE signal \cite{Huang:2012,Bougiatioti:2017,Meier:2013,Ramos:2013}. Moreover, other spin caloritronic and spintronic effects, such as the recently observed spin Hall magnetoresistance \cite{Nakayama:2013,Chen:2013} i.e., the magnetization orientation dependent absorption and reflection of a spin current density flowing along the direction normal to a FM insulator/normal metal (NM) interface, can benefit from the lack of mobile charge carriers in insulating materials, such as yttrium iron garnet (YIG). In conducting materials the anisotropic magnetoresistance can introduce a parasitic voltage hampering the correct appraisal of the spin Hall magnetoresistance signal \cite{Althammer:2013}. In order to disentangle the different contributions for different conductivity regimes, a material with controllable conductivity is required.\\
\mbox{ } Apart from YIG different insulating or semiconducting FMs from the group of spinel ferrites fulfill the requirements to be implemented in spin caloritronic devices. In this study, we focus on the fabrication and characterization of the spinel ferrite NiFe$_\textrm{2}$O$_\textrm{x}$. By reducing the oxygen content below its stoichiometric value (x=4), we vary the conductivity, the optical band gap and the electrical transport mechanisms. In particular, the high Curie temperatures (T$_\textrm{C}\approx 850\,$K) \cite{Pauthenet:1952,Dionne:1988} of spinel ferrites like NiFe$_\textrm{2}$O$_\textrm{4}$ (NFO) render them attractive candidates for future applications. Several investigations on the synthesis and properties of ferrite thin films have been reported so far. Pulsed laser deposition has been introduced as the most common deposition technique for high quality thin ferrite films \cite{Datta:2010,Caltun:2004,Hoppe:2015}. However, chemical vapor deposition \cite{Li:2011}, molecular beam epitaxy \cite{KuschelO:2016} and sputter deposition \cite{Klewe:2014} have also been used. Here, we fabricated high-quality epitaxial NiFe$_2$O$_\textrm{x}$ films by reactive magnetron co-sputter deposition varying the oxygen partial pressure during deposition. A major advantage of NFO compared to other spinel ferrites is attributed to the manipulation of the parasitic effects like the ANE by changing the base temperature of the measurement, due to its semiconducting character.\\
\mbox{ } So far, several investigations have been performed on FM/NM bilayers such as NFO/Pt in order to study the transport phenomena of those systems \cite{Bougiatioti:2017,Kuschel:2016,Kuschel:2015,MeierAIP:2016,Meier:2015,Meier:2013,Juan:2017} providing information about the spin Seebeck effect coefficient, the proximity induced magnetization in the Pt layer and the magnon spin transport in NFO. More specifically, the investigation of the temperature dependent electrical resistivity of the films leads to a deeper insight of the localization of electronic states and the disorder of such systems which influences the transport phenomena in those materials. In our systems, the temperature dependent resistivity of the films is usually explained in terms of band and hopping conduction. The thermal activation energy extracted from the resistivity measurements decreases progressively as the temperature falls indicating that, although the conduction mechanism at high temperatures is described by Arrhenius law (see Eq. (\ref{eq:resistivity_universal}) with $P=1$), at sufficiently low temperatures the conduction is controlled by the variable-range hopping mechanism.\\
\mbox{ } Except from the electrical properties of the films we additionally analyzed the optical properties by determining the optical band gap of the materials verifying their semiconducting origin which enables a profound understanding of the transport phenomena in those films \cite{Bougiatioti:2017}. Our results reveal states within the band gap which display charge transfer, creating a framework for realizing the electronic structure of those complex oxides that promotes their use in spintronics applications.

\section{Experimental techniques}

We fabricated NiFe$_2$O$_{\textrm{x}}$ ($4\geq x>0$) films by ultra high vacuum reactive dc sputter deposition from elemental Ni and Fe targets \cite{Klewe:2014} on MgAl$_2$O$_4$(001) (MAO) substrates with a lattice mismatch of about 3\%. Starting from pure high-resistive NFO $(\sim160\,$nm) and by changing the oxygen flow during deposition we fabricated NiFe$_2$O$_{\textrm{x}_1}$ (60$\,$nm) and NiFe$_2$O$_{\textrm{x}_2}$ ($35\,$nm) films, with $4>x_1>x_2>0$. The NFO was grown in pure O$_2$ atmosphere with a pressure of $2\cdot10^{-3\,}$mbar at 610\si{\degree}$\,$C substrate temperature. The NiFe$_2$O$_{\textrm{x}_{1}}$ and NiFe$_2$O$_{\textrm{x}_{2}}$ films were grown in Ar and O$_2$ atmosphere at 610\si{\degree}$\,$C substrate temperature. For the NiFe$_2$O$_{\textrm{x}_1}$ sample the Ar partial pressure during the deposition was $1.7\cdot10^{-3}\,$mbar, while the total pressure was $2\cdot10^{-3}$\,mbar. For the NiFe$_2$O$_{\textrm{x}_2}$ sample the Ar partial pressure was $1.8\cdot10^{-3}\,$mbar, while the total one was $2.3\cdot10^{-3}$\,mbar. The base pressure in all cases was less than 10$^{-8}$\,mbar. 
We additionally applied a continuous rotation of the substrate equal to 5 revolutions per minute during deposition in order to achieve a homogeneous growth of the films. The deficiency of oxygen in the NiFe$_2$O$_{\textrm{x}_1}$ and NiFe$_2$O$_{\textrm{x}_2}$ compared to NFO manifests itself via the room temperature (RT) electrical conductivity of the samples. The measured RT electrical conductivity value for the NFO, NiFe$_2$O$_{\textrm{x}_1}$ and NiFe$_2$O$_{\textrm{x}_2}$ sample is equal to 2.45$\,\cdot\,10^{-5}\,$kS/m, 6.67$\,$kS/m and 22.22$\,$kS/m, respectively.\\
\mbox{ } In order to determine the film thickness, x-ray reflectivity (XRR) measurements were performed in a Philips X’Pert Pro diffractometer in a Bragg-Brentano configuration, with a Cu K$_\mathrm{\alpha}$ source. In the same setup, x-ray diffraction (XRD) measurements were carried out using $\theta-2\theta$ and off-specular $\omega-2\theta$ scans for the study of the crystallographic properties of the films.\\
\mbox{ } The appropriate sputter parameters were adjusted after evaluating the x-ray fluorescence measurements to achieve the desired stoichiometry. The final Fe:Ni ratios were between 2.00 and 1.87, very close to the correct stoichiometric compositions. The oxygen content could not be derived quantitatively due to the insensitivity of the fluorescence detector regarding oxygen.\\
\mbox{ } The magnetic properties of the films were investigated by alternating gradient magnetometry (AGM) in a Princeton MicroMag using a magnetic field up to 1.3$\,$T. Furthermore, the electrical properties were examined by performing temperature dependent dc resistivity measurements using a two-point probe technique in a cryostat.\\
\mbox{ } The semiconductor type of the samples was extracted via Hall effect measurements at RT. The measurements were performed in a closed cycle helium cryostat by Cryogenic with magnetic fields up to 4$\,$T.\\
\mbox{ } The optical properties were investigated via ultraviolet-visible ($\textit{UV-Vis}$) spectroscopy in the range of 1.0 to 4.1$\,$eV (1200-300$\,$nm) in a Perkin Elmer Lambda 950 Spectrometer. Both reflection and transmission spectra were recorded in order to extract the absorption coefficient and derive the optical band gap energies.

\begin{figure}[!ht]
    \centering
    \includegraphics[height=8cm, width=\linewidth]{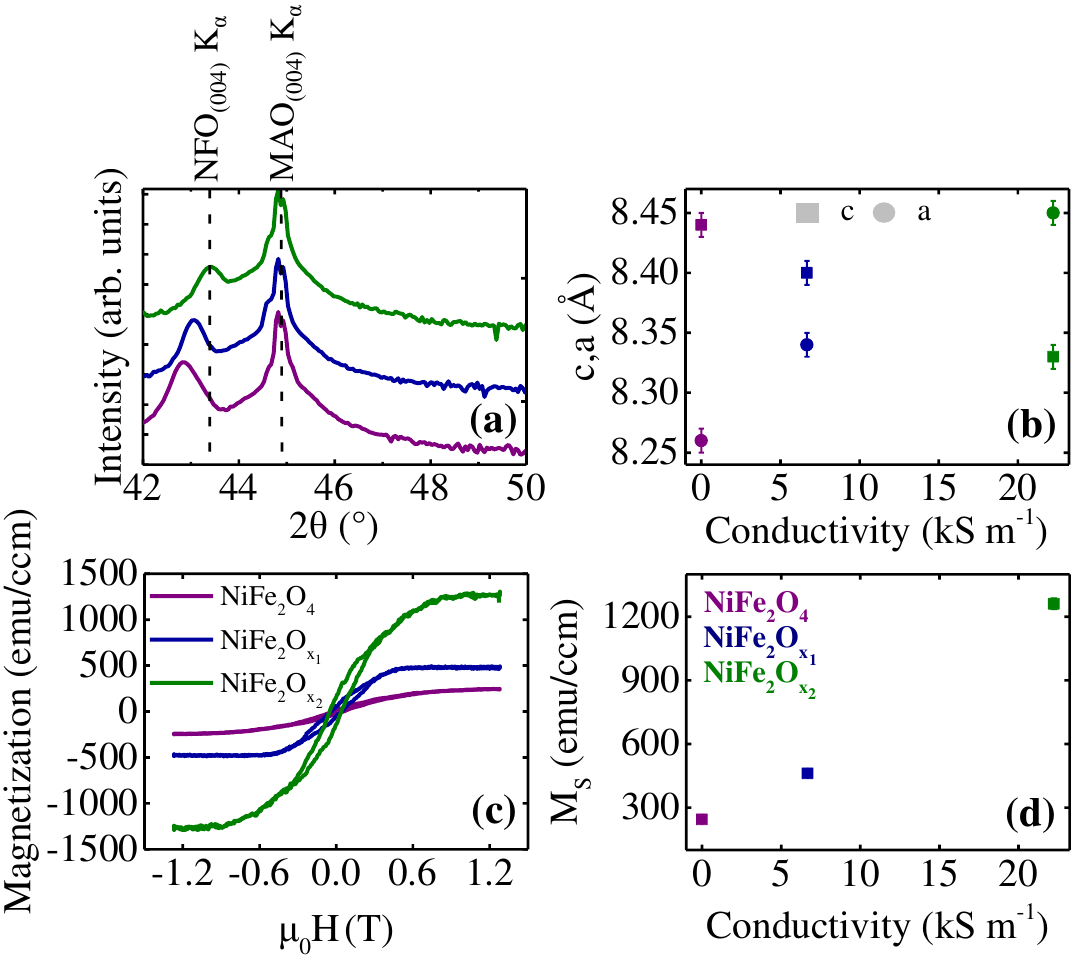}
  \caption{(a) XRD patterns for NiFe$_2$O$_{\textrm{4}}$, NiFe$_2$O$_{\textrm{x}_1}$ and NiFe$_2$O$_{\textrm{x}_2}$ samples. (b) Out-of-plane, $c$, and in-plane, \textsl{a}, lattice parameters plotted against the conductivity at RT for all films. (c) Magnetization curves collected via AGM. A linear diamagnetic background was subtracted. (d) Saturation magnetization, $M_{\textrm{S}}$, acquired at 1.3$\,$T plotted against the conductivity.}
    \label{fig:XRD_AGM}
\end{figure}

\section{Structural and magnetic properties}

Figure \ref{fig:XRD_AGM}(a) illustrates the results of $\theta-2\theta$ measurements for all samples. In the x-ray diffraction patterns, (004) Bragg peaks are visible for all samples showing a crystalline structure with epitaxial growth in [001] direction. From the peak positions in the XRD patterns the out-of-plane lattice constants, $c$, can be derived. Additionally, the in-plane lattice constants, $a$, can be identified from the position of the (606)-Peak ($2\theta\approx 103^\circ$), which is observable with off-specular $\omega-2\theta$ measurements and an $\omega$-offset of $\Delta\omega\approx 45^\circ$ (this is analog to an eccentric tilt of the sample around the [010] MAO direction by $\Delta\omega $). The obtained out-of-plane and in-plane lattice constants are presented in Fig. \ref{fig:XRD_AGM}(b) as a function of the conductivity for all samples.\\
\mbox{ } For NFO a tetragonal distortion is visible which is in agreement with epitaxial strain due to the lattice mismatch between NFO and the MAO substrate, since $\textsl{a}_\textrm{MAO}=8.08\,$\r{A}. Specifically, the film is expanded in the direction perpendicular to the surface ($c_{\textrm{NFO}}>c_{\textrm{bulk}}$) and compressed in the film plane ($\textsl{a}_{\textrm{NFO}}<\textsl{a}_{\textrm{bulk}}$). The bulk lattice constant value for NFO equal to $\textsl{a}_{\textrm{bulk}}=8.34\,$\r{A} is taken from Ref. \cite{Li:1991}. In order to quantify the strain effect in the film, the Poisson ratio is commonly used. From the formula \cite{Harrison:2005}

\vspace{-1em}
\begin{equation} \label{eq:Poisson}
\nu = -\frac{\epsilon_{\mathrm{oop}}}{\epsilon_{\mathrm{ip}}}=\frac{(c-\textrm{a}_{\mathrm{bulk}})}{(\textrm{a}-\textrm{a}_{\mathrm{bulk}})}
\end{equation}
where $\epsilon_{\mathrm{oop}}$ is the out-of-plane strain and $\epsilon_{\mathrm{ip}}$ is the in-plane strain, we extracted a positive strain equal to $\nu=1.25$. This value comes in line with Fritsch and Ederer \cite{Fritsch:2010} who reported a value of $\nu\approx1.2$ for NFO with in-plane compressive strain. The unit cell volume is reduced by about 1$\%$ with respect to bulk material.
For the other two NiFe$_\textrm{2}$O$_\textrm{x}$ films the in-plane lattice constants increase and the out-of-plane ones decrease compared to NFO. However, no further conclusion can be drawn for these samples since the corresponding bulk values of the lattice constants are not known and may differ from the ones of the NFO.\\
\mbox{ } Moreover, by performing $\omega$-scans around the (004) Bragg peak of our NiFe$_2$O$_{\textrm{x}}$ samples (not shown) we obtained a full width of half maximum equal to 1.2$\,^{\circ}$, 0.6$\,^{\circ}$, 0.7$\,^{\circ}$ for NFO, NiFe$_2$O$_{\textrm{x}_1}$ and NiFe$_2$O$_{\textrm{x}_2}$ samples, respectively. These values are slightly higher but still comparable to previous publications on NFO films prepared by pulsed laser deposition \cite{Ma:2010}.\\
\mbox{ } Figure \ref{fig:XRD_AGM}(c) illustrates the magnetization plotted against the magnetic field extracted from the AGM measurements. The plots are presented after the subtraction of diamagnetic contributions. In the field of 1.3$\,$T the NiFe$_2$O$_{\textrm{x}_1}$ and NiFe$_2$O$_{\textrm{x}_2}$ samples are clearly saturated. In contrast, for the NFO sample we reached 88$\%$ of saturation in the applied magnetic field. The saturation value was estimated from the anomalous Hall effect measurements with the application of a field strength equal to 4$\,$T. Figure \ref{fig:XRD_AGM}(d) shows the saturation magnetization values $M_{\textrm{S}}$ as a function of the conductivity for all samples. For the NFO the $M_{\textrm{S}}$ value equal to 244$\,$emu/ccm (in 88$\,\%$ saturation state) is consistent to earlier publications \cite{Klewe:2014}. Moreover, it is clearly observed that the magnetization increases with the increase in conductivity. One possible explanation for the increased magnetization is a higher ratio between magnetic ions and the non-magnetic oxygen in the lattice. This increased Fe and Ni density enables a larger moment per f.u. Additionally, deviations from the default NiFe$_2$O$_{\textrm{4}}$ stoichiometry potentially reduce the antiferromagnetic coupling between tetrahedral and octahedral lattice sites, leading to a larger net moment.

\section{Electrical properties}

\begin{figure}[!ht]
    \centering
    \includegraphics[height=4.1cm, width=\linewidth]{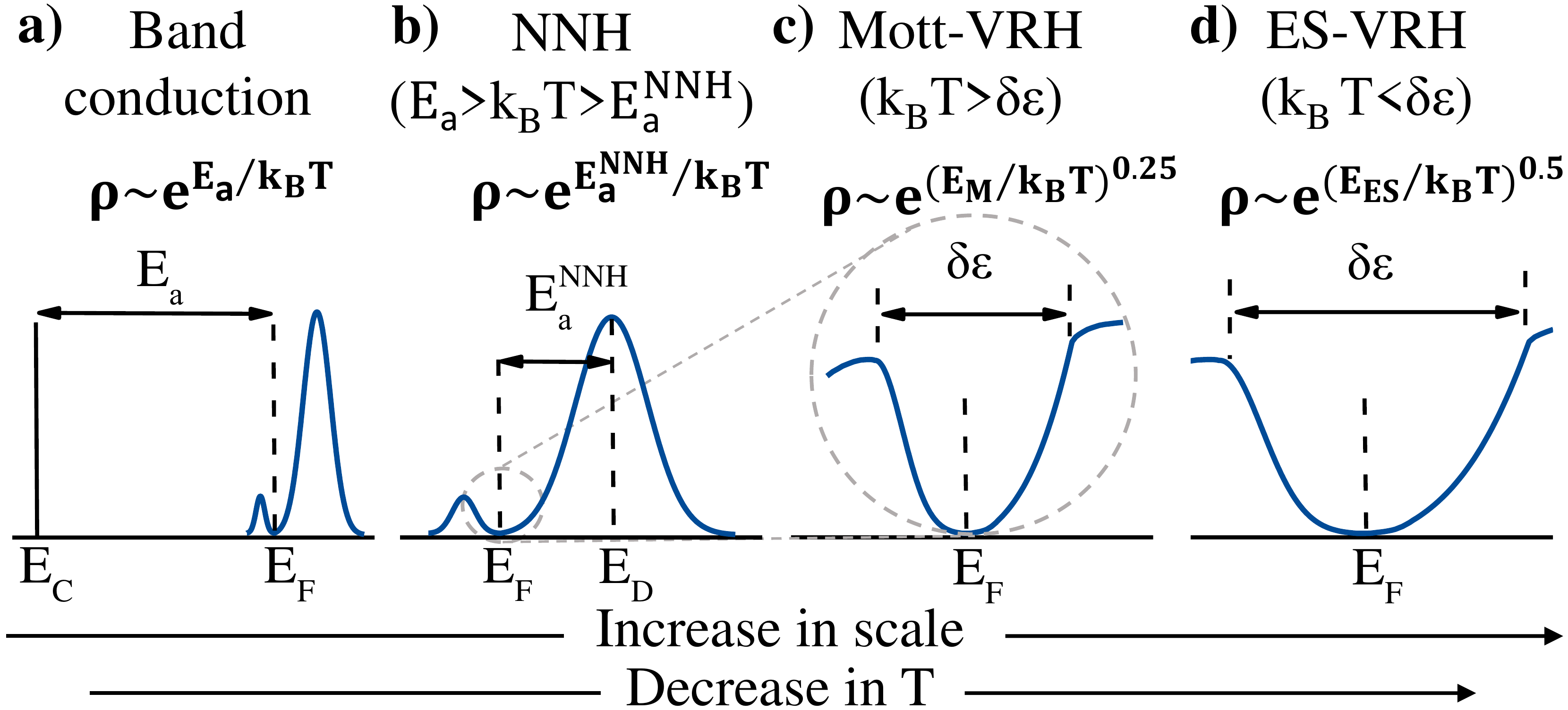}
  \caption{Conductivity mechanisms as a function of the temperature decrease representing qualitatively the DOS in a lightly doped semiconductor. (a) Band conduction. (b) Nearest-Neighbour hopping (NNH). (c) Mott-Variable Range Hopping (Mott-VRH). (d) Efros-Shklovskii-Variable Range Hopping (ES-VRH).}
    \label{fig:mechanisms}
\end{figure}

\begin{table}[ht]
\caption {Summary of different conduction mechanisms which take place in semiconductors with the corresponding characteristic energy $E_\textrm{t}$ and the value of the exponent $P$ from Eq. $(\ref{eq:resistivity_universal})$.}
\begin{center}
\begin{tabular}{ c|c|c }
\hline
\hline
Conduction mechanism & Characteristic energy $E_\textrm{t}$  & Exponent $P$  \rule{0pt}{2.5ex} \\
\hline
 
 Band conduction  & $E_\mathrm{a}$   & 1 \\
 NNH  &  $E_\mathrm{a}^\textrm{NNH}<E_\mathrm{a}$  & 1  \\
 Mott-VRH & $E_\textrm{M}=\textrm{k}_\textrm{B}\,\cdot\,\textrm{T}_\textrm{M}$  & 0.25 \\
 ES-VRH & $E_\textrm{ES}=\textrm{k}_\textrm{B}\,\cdot\,\textrm{T}_\textrm{ES}$  & 0.50 \\

\hline
\hline

\end{tabular}
\label{table:0.5}
\end{center}
\end{table}
In a generalized picture the electrical conduction in semiconductors consists of two types, the band and hopping conduction. Figure \ref{fig:mechanisms} represents the sequence of conductivity mechanisms replacing one by another as a function of the temperature decrease in a lightly doped (n-type) semiconductor \cite{Gantmakher:2005}. This sequence includes the assumption that the Fermi level, $E_\textrm{F}$, is located in the impurity band of the localized states of a doped semiconductor. A lower temperature value is balanced with a smaller section of the energy scale to distinguish between the different transitions of the carriers for the corresponding applied energy k$_\textrm{B}$T. In band conduction, charge carriers from localized states are thermally activated and transported to delocalized states, as visible in Fig. \ref{fig:mechanisms}(a). The highest energy at which states are still localized defines a mobility edge. The universal equation which describes the temperature dependent electrical resistivity in semiconductors is given by

\vspace{-1em}
\begin{equation} \label{eq:resistivity_universal}
\rho(E) = \rho_\textrm{0}\,\textrm{exp} \bigg[ \Big(\frac{E_\textrm{t}}{k_\textrm{B}T}\Big)^P \bigg]
\end{equation}
where $\rho_\textrm{0}$ is the resistivity at $T\,\to\,\infty $, $E_{\textrm{t}}$ is the transition energy, $k_\textrm{B}$ is the Boltzmann constant and $P\,(\,>\,0)$ is the characteristic exponent. The value of the exponent $P$ distinguishes between different conduction mechanisms by expressing the profile of the density of states (DOS). In band conduction, $P$=1 and $E_{\textrm{t}}$ corresponds to the thermal activation energy for the delocalization of carriers $E_\mathrm{a}$, as summarized in Table \ref{table:0.5}. $E_{\textrm{t}}$ is given by either $E_\textrm{C}-E_\textrm{F}$ or $E_\textrm{F}-E_\textrm{V}$, depending on whether electrons or holes are the charge carriers of the material. $E_\textrm{C}$, $E_\textrm{V}$ and $E_\textrm{F}$ are the mobility edges of the conduction band, the valence band and the Fermi energy, respectively.\\
\mbox{ } In hopping conduction, charge is transported through localized states in the vicinity of the Fermi energy (cf. Figs. \ref{fig:mechanisms}(b-d)). The conductivity is defined by electrons hopping directly between localized states in the impurity band without any excitation to the conduction band since they have insufficient energy for this transition. Therefore, the free electron band conduction is less important in this case \cite{ES:1984}. In this regime, there are two types of conduction mechanisms, the Nearest-Neighbour Hopping (NNH) and the Variable Range Hopping (VRH). In NNH, the hopping conductivity is expressed by transitions between the nearest neighbours (Fig. \ref{fig:mechanisms}(b)). The DOS in the donor impurity band at low donor concentration is maximum when the energy is of the order of the ionization energy of an isolated donor, $E_{\textrm{D}}$. When the initial and final states of such a transition are among the nearest neighbours, it is most probable that the corresponding energy levels are in the vicinity of the maximum DOS. The necessary condition for the NNH conduction to occur is the existence of a large number of pairs of close neighbour states with one of them being free. The  corresponding probability of this free state (for an n-type semiconductor) depends on its energy with respect to the Fermi level and is proportional to

\vspace{-1em}
\begin{equation} \label{eq:probability}
\textrm{exp}\bigg(\frac{-\lvert E_\textrm{F}-E_\textrm{D}\rvert}{k_\textrm{B}\,T}\bigg)\ \ \ . 
\end{equation}
Then, from the general semiconductor equation (Eq. (\ref{eq:resistivity_universal})) $E_{\textrm{t}}$ is now symbolized as $E_\textrm{a}^\textrm{NNH}$ and corresponds to the thermal activation energy having a smaller value compared to the energy required for thermally activated band conduction ($E_\mathrm{a}$), as summarized in Table \ref{table:0.5}.\\
\mbox{ } In relatively low disordered systems the further decrease of temperature such that $k_\textrm{B}T<<\lvert E_\textrm{F}-E_\textrm{D}\rvert$ causes the number of empty states among the nearest neighbours to be significantly small and, therefore, the electron hopping will take place between free states localized in the vicinity of the Fermi level symbolized by $\delta \varepsilon$. The average hopping length depends on temperature and the conduction mechanism changes now from NNH to VRH. When the VRH dominates the conduction, the condition $0<P<1$ for the exponent $P$ in Eq. (\ref{eq:resistivity_universal}) is fulfilled. The VRH model was firstly proposed by Mott \cite{Mott:1969} when he considered a constant density of states $N(E)$ near the Fermi level, supporting that the Coulomb interaction of electrons is weak and can be neglected. Thus, he showed that the value of the exponent $P$ in Eq. (\ref{eq:resistivity_universal}) is equal to 0.25 and the transition energy $E_{\textrm{t}}$ is given by $E_{\textrm{M}}\,=\,k_{\textrm{B}}T_{\textrm{M}}$ (cf. Fig. \ref{fig:mechanisms}(c)), where $E_{\textrm{M}}$ is the energy that corresponds to the characteristic Mott temperature $T_\textrm{M}$, as described in Table \ref{table:0.5}. $T_\textrm{M}$ can be correlated to the localization length $L_\textrm{c}$ via the formula

\vspace{-1em}
\begin{equation} \label{eq:temperature_Mott}
T_\textrm{M} = \frac{18}{L_\textrm{c}^3\,N(E_\textrm{F})\,k_\textrm{B}}
\end{equation}
where $N(E_\textrm{F})$ is the DOS at the Fermi level.\par
However, Efros and Shklovskii \cite{ES:1984} later on suggested that at low enough temperatures for highly disordered systems, $N(E)$ can not be considered constant anymore, but behaves as $N(E)\propto(E-E_\textrm{F})^2$. This behaviour derives from the energetic insufficiency of the system to overcome the electron-hole Coulomb interaction arising from the movement of the electron from one state to the other. This vanishing DOS is called Coulomb gap. In the Efros-Shklovskii-VRH (ES-VRH) conduction regime (Fig. \ref{fig:mechanisms}(d)) the exponent $P$ is equal to 0.5. Moreover, $E_{\textrm{t}}$ is now given by $E_{\textrm{ES}}\,=\,k_{\textrm{B}}T_{\textrm{ES}}$, where $E_{\textrm{ES}}$ is the energy that corresponds to the characteristic Efros-Shklovskii temperature $T_\textrm{ES}$, as included in Table \ref{table:0.5}. $T_\textrm{ES}$ is given by

\vspace{-1em}
\begin{equation} \label{eq:temperature_ES}
T_\textrm{ES} = \frac{2.8\,e^2}{\epsilon\,L_\textrm{c}\,k_\textrm{B}}
\end{equation}
where $\epsilon$ is the dielectric constant and $e$ is the electron charge. At intermediate disordered systems, there may be a crossover from ES- to Mott-VRH with increasing temperature.

\begin{figure}[!ht]
    \centering
    \includegraphics[height=7cm, width=\linewidth]{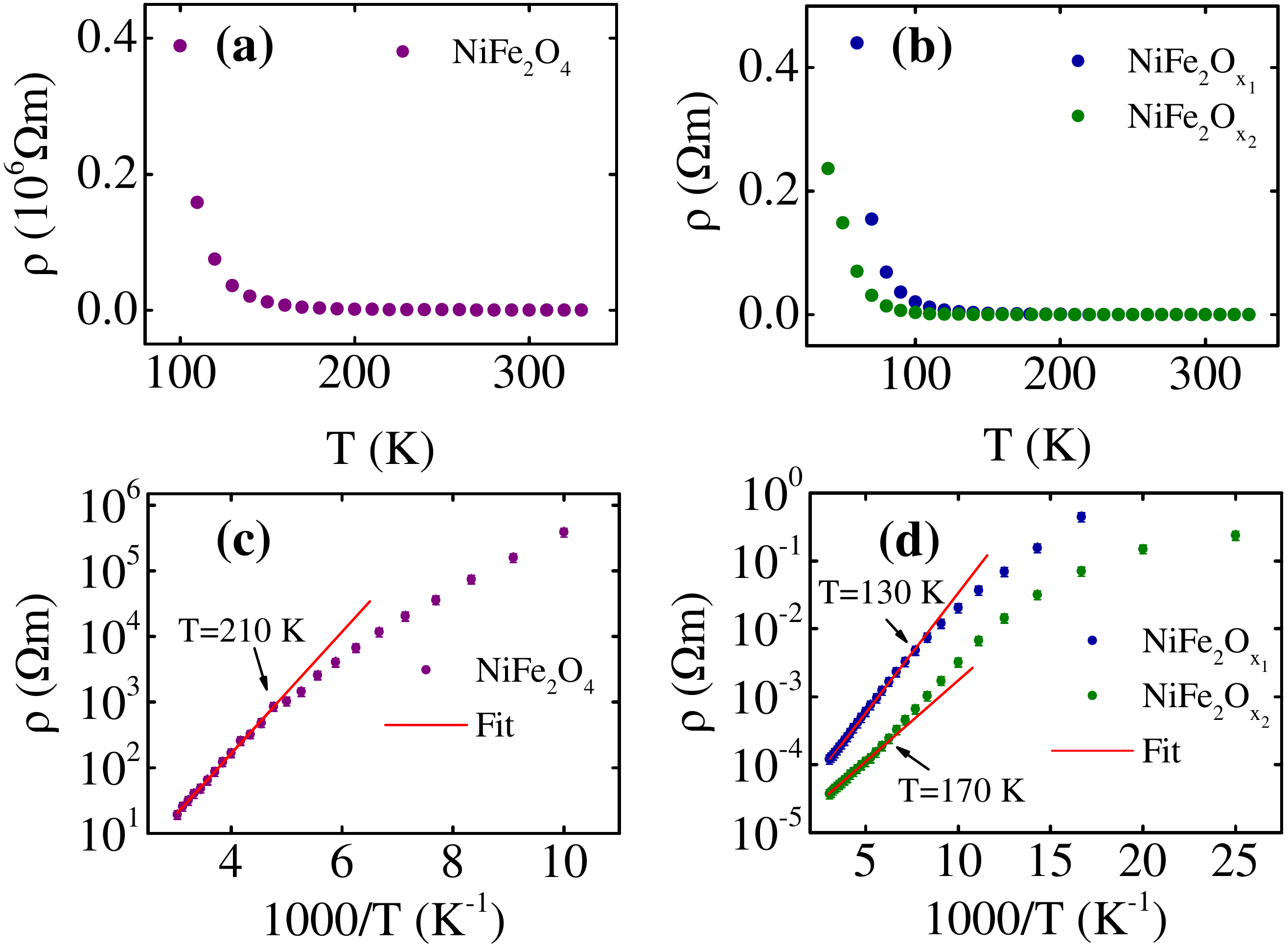}
  \caption{(a),(b) Temperature dependence of the resistivity and (c),(d) Arrhenius plots for NiFe$_2$O$_{\textrm{4}}$, NiFe$_2$O$_{\textrm{x}_1}$ and NiFe$_2$O$_{\textrm{x}_2}$ samples, respectively. In the high temperature regime the straight line segments fit the data closely indicating the validity of Arrhenius law. The black arrows note the lowest temperature point included in the linear fit.}
    \label{fig:Arrhenius}
\end{figure}

\begin{figure}[!ht]
    \centering
    \includegraphics[height=6.5cm, width=\linewidth]{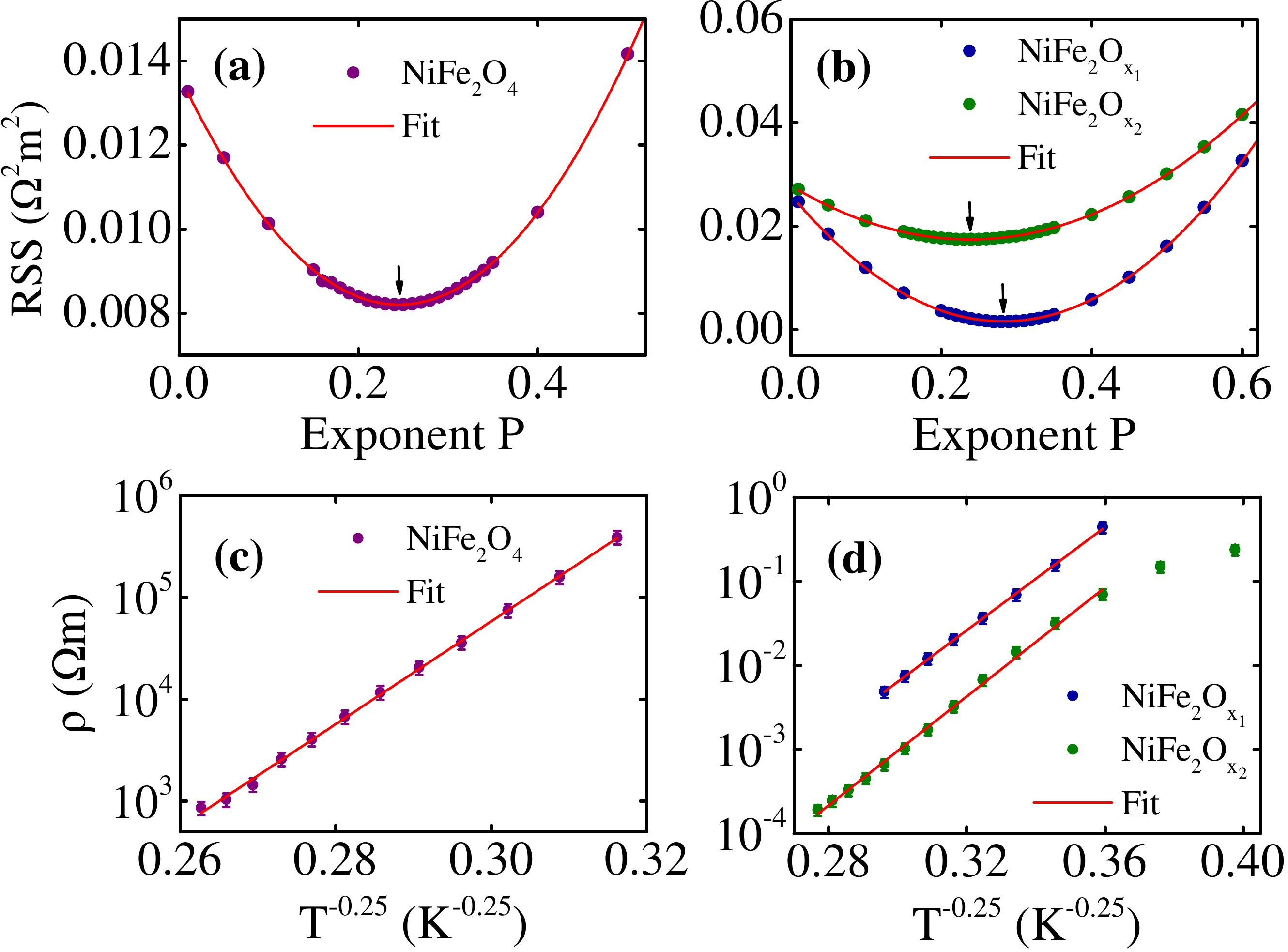}
  \caption{(a),(b) Plot of the residual sum of squares (RSS) against the exponent $P$ from Eq. (\ref{eq:resistivity_universal}) fitted with parabolas for NiFe$_2$O$_{\textrm{4}}$, NiFe$_2$O$_{\textrm{x}_1}$ and NiFe$_2$O$_{\textrm{x}_2}$ samples. The black arrows denote the minimum of the parabola fit indicating which conduction mechanism governs the measurements in the low temperature regimes. (c),(d) Temperature dependence of the resistivity in the Mott-VRH regime.}
    \label{fig:resistivity_Mott}
\end{figure}

In our systems from highly resistive to semiconducting-like NiFe$_\textrm{2}$O$_\textrm{x}$ we expect to observe different conduction mechanisms that contribute in the examined temperature range. Figures \ref{fig:Arrhenius}(a) and (b) illustrate the electrical resistivity $\rho$ as a function of temperature ranging between (100$\,-\,$330)$\,\textrm{K}$ for NFO, (60$\,-\,$330)$\,\textrm{K}$ for NiFe$_2$O$_{\textrm{x}_1}$ and (40$\,-\,$330)$\,\textrm{K}$ for NiFe$_2$O$_{\textrm{x}_2}$. The expected semiconducting behaviour with increasing resistivity for decreasing temperature is clearly observed in all cases. In order to investigate the conduction mechanisms governing our systems we firstly considered the simplest form of thermal activation process. The temperature dependent resistivity can be described by Arrhenius law which corresponds to $P=1$ in Eq. (\ref{eq:resistivity_universal}), as included in Table \ref{table:0.5}. The required energy for the thermally activated charge transport can be derived by a linear regression of the temperature dependent resistivity. Figures \ref{fig:Arrhenius} (c) and (d) show the Arrhenius plot of ln($\rho$) vs. 1000/$T$ for all samples. The experimental data were fitted with Eq. (\ref{eq:resistivity_universal}) for $P$=1 in order to determine the thermal activation energy. In the high temperature regime the straight line segments fit the data closely. The black arrows indicate the lowest temperature point included in the linear fit. Consequently, the thermal activation energy value for NFO was found to be equal to $E_\mathrm{a}^\textrm{NFO}=0.19\,\textrm{eV}$. This result is in accordance with our previous investigations on sputter-deposited and chemical vapor deposited NFO \cite{Klewe:2014,Meier:2013}, as well as the values found by Lord and Parker in sintered NFO specimens, Austin and Elwell in NFO single crystals, and Ponpandian \textrm{et al}. in NFO nanoparticles \cite{Lord:1960,Austin:1970,Ponpandian:2002}. For the NiFe$_2$O$_{\textrm{x}_1}$ and NiFe$_2$O$_{\textrm{x}_2}$ films we found $E_\mathrm{a}^{\textrm{NiFe}_2\textrm{O}_{\textrm{x}_1}}=0.07\,\textrm{eV}$ and $E_\mathrm{a}^{\textrm{NiFe}_2\textrm{O}_{\textrm{x}_2}}=0.05\,\textrm{eV}$, respectively. The thermal activation energy is lower in the more conducting samples reflecting the  additional electronic states localized in the band gap which contribute to the measured resistivity.\\
\mbox{ } On the contrary, in the low temperature regime the plots show significant deviations from the straight lines (as visible in Figs. \ref{fig:Arrhenius} (c) and (d)) suggesting that the conduction mechanism in which the carriers are thermally activated and jump over a certain semiconductor energy barrier cannot be the dominant one and a crossover between two mechanisms is reasonable. In order to determine with sufficient accuracy which conduction mechanism governs the resistivity we plotted the data in low temperature regimes ln($\rho$) vs. $T^{\textrm{-P}}$ where we varied the exponent $P$ from 0.1 to 1 with outer steps of 0.05 and inner steps of 0.01 near the minimum. We fitted the data by straight line segments and we plotted the residual sum of squares (RSS) versus the exponent $P$. From the parabola fits we extracted the appropriate exponent $P$ which minimizes the RSS leading to the best fitting.\\
\mbox{ } Figures \ref{fig:resistivity_Mott}(a) and (b) illustrate the RSS as a function of the exponent $P$ in the low temperature regimes, (100$\,-\,$210)$\,$K for NFO, (60$\,-\,$130)$\,$K for NiFe$_2$O$_{\textrm{x}_1}$ and (60$\,-\,$170)$\,$K for NiFe$_2$O$_{\textrm{x}_2}$. All curves were fitted with parabolas to estimate the minimum of the corresponding curve with high precision. The black arrows indicate the minimum of the parabola fit. We found that the minimum of the RSS is $P$=0.25 for NFO, $P$=0.28 for NiFe$_2$O$_{\textrm{x}_1}$ and $P$=0.24 for NiFe$_2$O$_{\textrm{x}_2}$. Thus, we deduce that the values of $P$ are 0.25 or very close to it indicating the existence of Mott-VRH conduction in the low temperature regimes. This suggests an almost constant DOS near the Fermi level. Figures \ref{fig:resistivity_Mott}(c) and (d) show ln($\rho$) plotted against $T^{-0.25}$, according to the Mott model, with the corresponding linear fits indicating that Mott-VRH model describes accurately the data in the low temperature regimes. The characteristic Mott temperature $T_\textrm{M}$ is extracted from Eq. $(\ref{eq:resistivity_universal})$, with $P$=0.25. For the NiFe$_2$O$_{\textrm{x}_2}$ sample the two last points at 50$\,$K and 40$\,$K were left out from the fitting since those points present a second change in the slope of the curve in the low temperature dependent resistivity. This behaviour could also indicate a possible transition between Mott-VRH and ES-VRH at even lower temperatures. However, this assumption requires further investigation in lower temperature ranges. The values of $\rho$ at RT, $E_\textrm{a}$ and $T_\textrm{M}$ are summarized in Table \ref{table:1} for all samples. The Mott temperature $T_\textrm{M}$ increases with increasing resistivity, as reported in Table \ref{table:1}, indicating that the quantity $N(E_\textrm{F})L_\textrm{c}^3$ (see Eq. \ref{eq:temperature_Mott}) is smaller. This behaviour is expected since in the most resistive samples $N(E_\textrm{F})$ has a smaller value as well.
\begin{table}[ht]
\caption {Resistivity $\rho$ collected at RT, thermal activation energy $E_\mathrm{a}$ and characteristic Mott temperature $T_\textrm{M}$ extracted from the linear fits in Figs. \ref{fig:Arrhenius} and \ref{fig:resistivity_Mott}.}
\begin{center}
\begin{tabular}{ c|c|c|c }

\hline
\hline
Film & $\rho\,(\Omega \cdot \textrm{m})$  & $E_\mathrm{a} \,(\textrm{eV})$ & $T_\textrm{M}\,(\textrm{K})$\\
\hline
 
NiFe$_2$O$_{\textrm{4}}$  & 40.5   & 0.19    & 5264 \\
NiFe$_2$O$_{\textrm{x}_1}$   &  $1.5\cdot10^{-4}$  & 0.07  & 2482 \\
NiFe$_2$O$_{\textrm{x}_2}$ & $4.5\cdot10^{-5}$  & 0.05 & 2400 \\

\hline
\hline

\end{tabular}
\label{table:1}
\end{center}
\end{table}

\section{Hall effect measurements}

\begin{figure}[!ht]
    \centering
    \includegraphics[height=3.5cm, width=\linewidth]{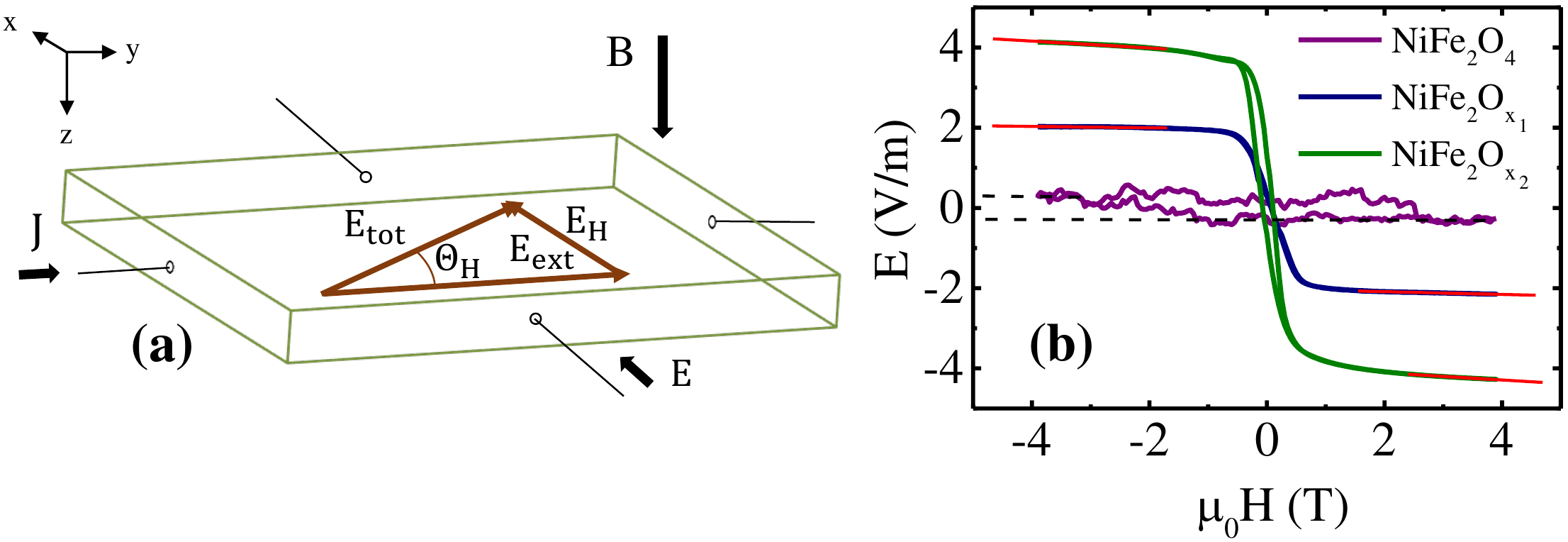}
  \caption{(a) Schematic illustration of Hall effect geometry. (b) Hall effect measurements for NiFe$_2$O$_{\textrm{4}}$, NiFe$_2$O$_{\textrm{x}_1}$ and NiFe$_2$O$_{\textrm{x}_2}$ samples. From the slope of the fitting curves the Hall coefficient is extracted.}
    \label{fig:HE}
\end{figure}

In order to extract the semiconductor type of our films we performed Hall effect measurements according to the geometry displayed in Fig. \ref{fig:HE}(a). A charge current was flowing along the y-axis in the presence of an out-of-plane magnetic field and the voltage along the x-axis, as well as the voltage along the y-axis, were recorded. Figure \ref{fig:HE}(b) displays the detected electric field, $E$, along the x-axis plotted against the applied external magnetic field for the NFO, NiFe$_2$O$_{\textrm{x}_1}$ and NiFe$_2$O$_{\textrm{x}_2}$ samples. Both curves for NiFe$_2$O$_{\textrm{x}_1}$ and NiFe$_2$O$_{\textrm{x}_2}$ show a slope changing linearly with the field in the region where the magnetization dependent anomalous Hall effect (AHE) saturates. This can be attributed to the ordinary Hall effect (OHE). From the slopes of the linear fits in the positive and negative saturation regimes of the curves, the ordinary Hall coefficient $R_\textrm{H}$ can be extracted by using

\vspace{-1em}
\begin{equation} \label{eq:R_Hall_general}
\textbf{E}_\textrm{H} = R_\textrm{H}(\textbf{J}\times \textbf{B})
\end{equation}
where $\textbf{E}_\textrm{H}$ is the OHE field, $\textbf{J}$ is the charge current density and $\textbf{B}$ is the external magnetic field. Considering the geometry displayed in Fig. \ref{fig:HE}(a) the Hall coefficient is given by

\vspace{-1em}
\begin{equation} \label{eq:R_Hall_general_2}
R_\textrm{H} = \frac{E_\textrm{H,x}}{J_\textrm{y}B_\textrm{z}}
\end{equation}
and taking into account the slope of the linear fits $A_\textrm{OHE}$ extracted from the curves, $R_\textrm{H}$ is modified to

\vspace{-1em}
\begin{equation} \label{eq:R_Hall}
R_\textrm{H} = \frac{A_\textrm{OHE}}{J_\textrm{y}}\ \ \ . 
\end{equation}

Considering a charge current density of $J_\textrm{y}\,$=$\,$1.04$\,\cdot\,10^7\,$A/m$^2$ (17.86$\,\cdot\,10^7\,$A/m$^2$) and a slope of $A_\textrm{OHE}\,$=$\,-0.0256\,$V/Tm ($A_\textrm{OHE}\,$=$\,-0.0870\,$V/Tm) for the NiFe$_2$O$_{\textrm{x}_1}$ (NiFe$_2$O$_{\textrm{x}_2}$) sample, we can extract the corresponding Hall coefficients. We find, $R_{\textrm{H}}^{\textrm{NiFe}_2\textrm{O}_{\textrm{x}_1}}\,=-$24$\,\cdot10^{-10}\,$m$^3$/C and $R_{\textrm{H}}^{\textrm{NiFe}_2\textrm{O}_{\textrm{x}_2}}\,$=$\,-4.9\cdot10^{-10}\,$m$^3$/C for the NiFe$_2$O$_{\textrm{x}_1}$ and NiFe$_2$O$_{\textrm{x}_2}$ samples, respectively. The large resistivity of the NFO makes the measurements challenging and prevents us to reliably determine the Hall coefficient and the semiconducting behaviour for this sample (see Fig. \ref{fig:HE}(b)).\\
\mbox{ } The extracted $R_\textrm{H}$ values are quite small and would indicate high charge carrier densities which are not representative for materials with semiconducting character. The impurity conduction originating from states above, at and below the Fermi level indicates that we cannot expect to have a pure n- or p-type semiconducting behaviour and, therefore, a mixed-type semiconducting behaviour could characterize our NiFe$_2$O$_{\textrm{x}_1}$ and NiFe$_2$O$_{\textrm{x}_2}$ samples. In a mixed-type of semiconductor both electrons and holes contribute and the Hall coefficient is given by the formula

\vspace{-1em}
\begin{equation} \label{eq:mixed}
R_\textrm{H} = \frac{R_\textrm{H,n}\sigma_\textrm{n}^{2}+R_\textrm{H,p}\sigma_\textrm{p}^{2}}{(\sigma_\textrm{n}+\sigma_\textrm{p})^2}
\end{equation}
where $\sigma_\textrm{n}$ is the conductivity of electrons, $\sigma_\textrm{p}$ is the conductivity of holes, $R_\textrm{H,n}$ is the Hall coefficient for electrons and $R_\textrm{H,p}$ is the Hall coefficient for holes. Since R$_\textrm{H,n}$ and R$_\textrm{H,p}$ have different signs, we could expect quite small values for the total R$_\textrm{H}$, like in our case.\\
\mbox{ } From the Hall effect geometry illustrated in Fig. \ref{fig:HE}(a) the two components  $E_\textrm{H}$ and $E_\textrm{ext}$ define a total electric field, $E_\textrm{tot}$. $E_\textrm{ext}$ is the external electric field along the y-axis and $E_\textrm{H}$ is the already investigated OHE field along the x-axis. Then the total electric field is tilted according to the Hall angle, $\Theta_\textrm{H}$. Consequently, the mobility of the carriers is obtained from the formula

\vspace{-1em}
\begin{equation} \label{eq:mixed_mobility}
\textrm{tan}\Theta_\textrm{H} = \frac{E_\textrm{H,x}}{E_\textrm{ext,y}}=\mu B_\textrm{z}
\end{equation}
where $\mu$ is the mobility of the carriers. Considering $B_\textrm{z}\,$=$\,4\,$T,  $E_\textrm{H,x}\,$=$\,0.10\,$V/m ($E_\textrm{H,x}\,$=$\,0.34\,$V/m) and $E_\textrm{ext,y}\,$=$\,97.31\,$V/m ($E_\textrm{ext,y}\,$=$\,43.13\,$V/m) for the NiFe$_2$O$_{\textrm{x}_1}$ (NiFe$_2$O$_{\textrm{x}_2}$) sample we extracted the corresponding mobilities. We find, $\mu_{\textrm{NiFe}_2\textrm{O}_{\textrm{x}_1}}$=$\,$2.6$\,$cm$^2$/Vs and $\mu_{\textrm{NiFe}_2\textrm{O}_{\textrm{x}_2}}$=$\,$19.7$\,$cm$^2$/Vs for the NiFe$_2$O$_{\textrm{x}_1}$ and NiFe$_2$O$_{\textrm{x}_2}$ samples, respectively. The NiFe$_2$O$_{\textrm{x}_2}$ has higher mobility value compared to the NiFe$_2$O$_{\textrm{x}_2}$ reflecting its more conducting character. The obtained mobility values for both samples are consistent with the ones reported for organic semiconductors used in thin film transistors \cite{Rolin:2017}.

\section{Optical properties}

\begin{figure}[!ht]
    \centering
    \includegraphics[height=6.5cm, width=\linewidth]{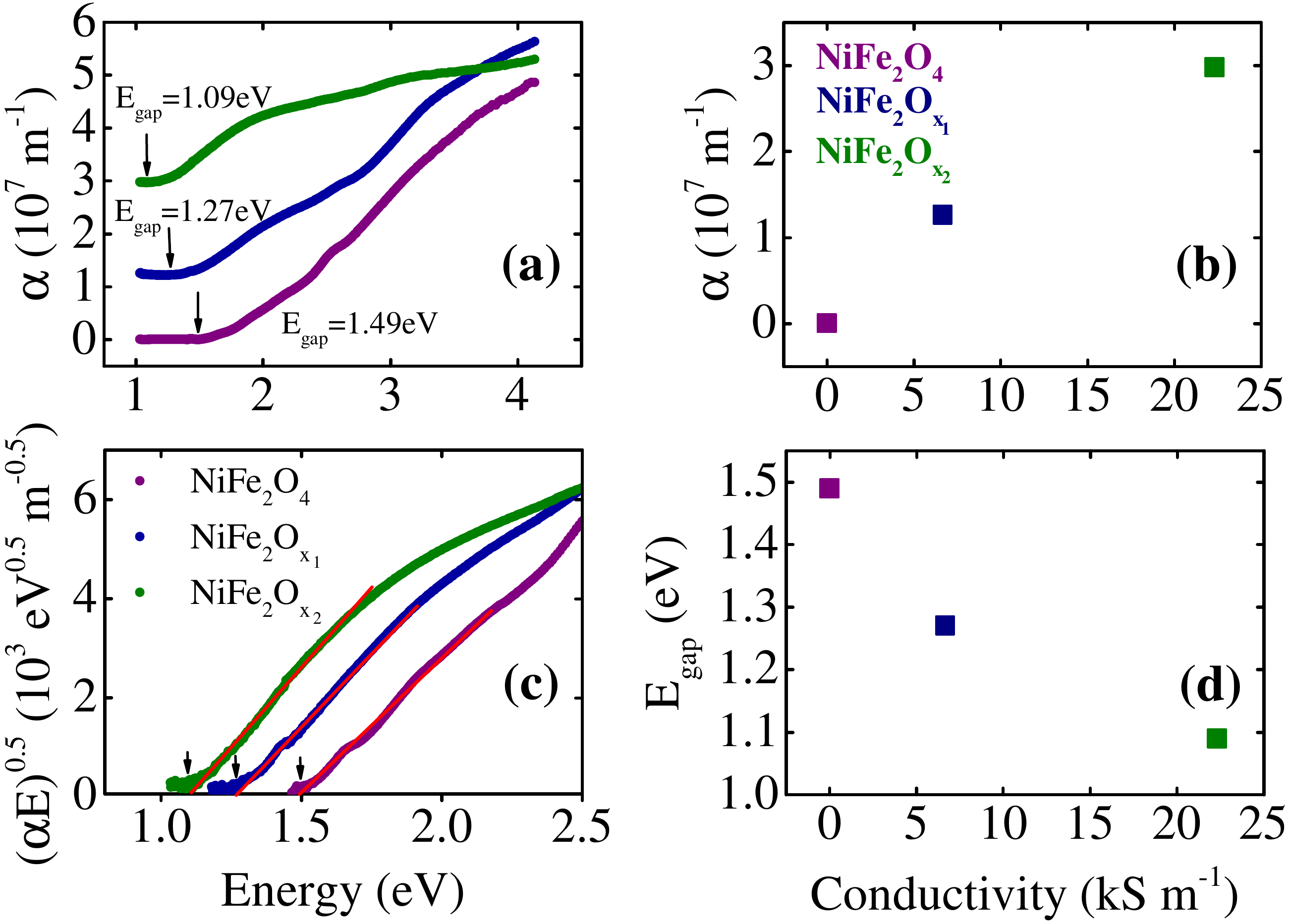}
  \caption{(a) The RT optical absorption spectrum for NFO, NiFe$_2$O$_{\textrm{x}_1}$ and NiFe$_2$O$_{\textrm{x}_2}$. (b) The absorption saturation value at 1$\,$eV versus the RT conductivity. (c) Tauc plot $(\alpha E)^{0.5}$ versus energy for the determination of the minimum gap for all samples. The black arrows indicate the minimum direct gaps. (d) The band gap energy versus the RT conductivity.}
    \label{fig:absorption_Tauc}
\end{figure}

To investigate the optical properties of the films the energy-dependent absorption coefficient $\alpha$(E) was extracted from the measured transmission $T$ and reflection $R$ spectra using

\vspace{-1em}
\begin{equation} \label{eq:absorption}
\alpha = \frac{1}{d}\textrm{ln}\frac{1-R}{T}
\end{equation}
where $d$ is the thickness of each sample. Figure \ref{fig:absorption_Tauc}(a) illustrates the absorption coefficient as a function of the energy for all samples. It is worth noting that down to a lower energy limit a saturation plateau is visible in all cases. The plateau region indicates that the energy of the photons is lower than the band gap of the sample and, therefore, insufficient to excite electrons from the valence band to the conduction band. Nevertheless, it is clearly observed that there is a different finite absorption value for each sample. Interestingly, as visible in  Fig. \ref{fig:absorption_Tauc}(b) where the absorption saturation value is plotted against the conductivity of the samples, for the NFO the absorption saturation value is zero in this plateau unveiling that the additional electronic states in the band gap are too few to noticeably contribute to the measured absorbance. On the other hand, for the NiFe$_2$O$_{\textrm{x}_1}$ and NiFe$_2$O$_{\textrm{x}_2}$ samples the absorption saturation value is equal to 1.26$\,\cdot\,$10$^7\,$m$^{-1}$ and 2.99$\,\cdot\,$10$^7\,$m$^{-1}$, respectively, confirming that the number of electronic states in the band gap contributing to the measured absorption is larger in the more conducting samples. The absorption spectrum of the NFO is very similar to our previous investigations \cite{Klewe:2014,Meinert:2014} as well as the epitaxial NFO from Holinsworth \textit{et al}. \cite{Holinsworth:2013}.\\
\mbox{ } A common way to determine the minimum gap from the optical absorption spectra is by evaluating Tauc plots as explained in Ref. \cite{Meinert:2014}. The indirect gap can be extracted from straight line segments in $(\alpha E)^{0.5}$ plotted over energy. In Fig. \ref{fig:absorption_Tauc}(c) the minimum gaps are extracted for each sample. It is worth mentioning that in order to extract the band gap energies the curves are shifted such that the plateau regions are at zero absorption. For the NFO the optical band gap is estimated to be $E_{\textrm{gap}}^{\textrm{NFO}} \approx 1.49\,\textrm{eV}$, close to previous publications \cite{Klewe:2014,Sun:2012}. The optical band gap for the NiFe$_2$O$_{\textrm{x}_1}$ and NiFe$_2$O$_{\textrm{x}_2}$ samples is $E_\textrm{gap}^{\textrm{NiFe}_2\textrm{O}_{\textrm{x}_1}} \approx 1.27\,\textrm{eV}$ and $E_\textrm{gap}^{\textrm{NiFe}_2\textrm{O}_{\textrm{x}_2}} \approx 1.09\,\textrm{eV}$, respectively, unveiling the more conducting character of the latter. The band gap energy of all samples is presented in Fig. \ref{fig:absorption_Tauc}(d) as a function of the RT electrical conductivity. However, the determination of the band gap using Tauc plots could lead to rough estimated values because of the uncertainties that may come up during this kind of data processing \cite{Meinert:2014}. It is clear that the band gap energy increases with the decrease of conductivity. In addition, the extracted optical band gap is considerably larger than the corresponding thermal activation energy estimated from the temperature dependent resistivity measurements, in all cases. The reason for that is focused on the sensitivity of the temperature dependent resistivity to all charge transport mechanisms which characterize the film, for example chemical impurities, defects etc, that influence the measured resistivity.

\section{Conclusion}

In conclusion, we fabricated NiFe$_\textrm{2}$O$_\textrm{x}$ samples on MgAl$_2$O$_4$ (001) substrates by reactive dc magnetron co-sputtering while varying the oxygen atmosphere during deposition in order to investigate the structural, magnetic, electronic and optical properties of the films. We extracted that the conduction mechanism in these systems differs in the examined low and high temperature regimes. Resistivity measurements in the high temperature regimes were fitted well with a model for Arrhenius type of conduction for the delocalization of the carriers allowing to extract the thermal activation energy of the samples. We obtained low thermal activation energies for all samples indicating a semiconducting behaviour. Furthermore, in the low temperature regimes Mott-VRH is the dominant conduction mechanism dictating that the electrical transport is supported by impurities in the crystal which hop between localized states in the impurity band. Moreover, using Hall effect measurements mixed-type semiconducting character was identified in the NiFe$_2$O$_{\textrm{x}_1}$ and NiFe$_2$O$_{\textrm{x}_2}$ films along with the corresponding mobilities of the charge carriers. From optical absorption spectra and the corresponding Tauc plots the optical band gap of the films were found to be significantly larger than the electrical ones which is in line with our previous investigations. The fabrication of a variable material with oxygen deficiency enables the manipulation of the electrical and optical properties, which is beneficial for the profound investigations of the transport phenomena and, therefore, forwards the implementation of such materials in spintronic and spin caloritronic applications. 

\section{Acknowledgement}

The authors gratefully acknowledge financial support by the Deutsche Forschungsgemeinschaft (DFG) within the Priority Program Spin Caloric Transport (SPP 1538). Further, they thank U. Heinzmann for making available the \textit{UV/Vis} spectrometer. C.K. acknowledges financial support
by the Alexander von Humboldt foundation.

\bibliographystyle{apsrev4-1}

\bibliography{main}

\begin{thebibliography}{39}%
\makeatletter
\providecommand \@ifxundefined [1]{%
 \@ifx{#1\undefined}
}%
\providecommand \@ifnum [1]{%
 \ifnum #1\expandafter \@firstoftwo
 \else \expandafter \@secondoftwo
 \fi
}%
\providecommand \@ifx [1]{%
 \ifx #1\expandafter \@firstoftwo
 \else \expandafter \@secondoftwo
 \fi
}%
\providecommand \natexlab [1]{#1}%
\providecommand \enquote  [1]{``#1''}%
\providecommand \bibnamefont  [1]{#1}%
\providecommand \bibfnamefont [1]{#1}%
\providecommand \citenamefont [1]{#1}%
\providecommand \href@noop [0]{\@secondoftwo}%
\providecommand \href [0]{\begingroup \@sanitize@url \@href}%
\providecommand \@href[1]{\@@startlink{#1}\@@href}%
\providecommand \@@href[1]{\endgroup#1\@@endlink}%
\providecommand \@sanitize@url [0]{\catcode `\\12\catcode `\$12\catcode
  `\&12\catcode `\#12\catcode `\^12\catcode `\_12\catcode `\%12\relax}%
\providecommand \@@startlink[1]{}%
\providecommand \@@endlink[0]{}%
\providecommand \url  [0]{\begingroup\@sanitize@url \@url }%
\providecommand \@url [1]{\endgroup\@href {#1}{\urlprefix }}%
\providecommand \urlprefix  [0]{URL }%
\providecommand \Eprint [0]{\href }%
\providecommand \doibase [0]{http://dx.doi.org/}%
\providecommand \selectlanguage [0]{\@gobble}%
\providecommand \bibinfo  [0]{\@secondoftwo}%
\providecommand \bibfield  [0]{\@secondoftwo}%
\providecommand \translation [1]{[#1]}%
\providecommand \BibitemOpen [0]{}%
\providecommand \bibitemStop [0]{}%
\providecommand \bibitemNoStop [0]{.\EOS\space}%
\providecommand \EOS [0]{\spacefactor3000\relax}%
\providecommand \BibitemShut  [1]{\csname bibitem#1\endcsname}%
\let\auto@bib@innerbib\@empty
\bibitem [{\citenamefont {Hoffmann}\ and\ \citenamefont
  {Bader}(2015)}]{Hoffmann:2015}%
  \BibitemOpen
  \bibfield  {author} {\bibinfo {author} {\bibfnamefont {A.}~\bibnamefont
  {Hoffmann}}\ and\ \bibinfo {author} {\bibfnamefont {S.~D.}\ \bibnamefont
  {Bader}},\ }\href {\doibase 10.1103/PhysRevApplied.4.047001} {\bibfield
  {journal} {\bibinfo  {journal} {Phys. Rev. Applied}\ }\textbf {\bibinfo
  {volume} {4}},\ \bibinfo {pages} {047001} (\bibinfo {year}
  {2015})}\BibitemShut {NoStop}%
\bibitem [{\citenamefont {Hirsch}(1999)}]{Hirsch:1999}%
  \BibitemOpen
  \bibfield  {author} {\bibinfo {author} {\bibfnamefont {J.~E.}\ \bibnamefont
  {Hirsch}},\ }\href {\doibase 10.1103/PhysRevLett.83.1834} {\bibfield
  {journal} {\bibinfo  {journal} {Phys. Rev. Lett.}\ }\textbf {\bibinfo
  {volume} {83}},\ \bibinfo {pages} {1834} (\bibinfo {year}
  {1999})}\BibitemShut {NoStop}%
\bibitem [{\citenamefont {Uchida}\ \emph {et~al.}(2008)\citenamefont {Uchida},
  \citenamefont {Takahashi}, \citenamefont {Harii}, \citenamefont {Ieda},
  \citenamefont {Koshibae}, \citenamefont {Ando}, \citenamefont {Maekawa},\
  and\ \citenamefont {Saitoh}}]{Uchida:2008}%
  \BibitemOpen
  \bibfield  {author} {\bibinfo {author} {\bibfnamefont {K.}~\bibnamefont
  {Uchida}}, \bibinfo {author} {\bibfnamefont {S.}~\bibnamefont {Takahashi}},
  \bibinfo {author} {\bibfnamefont {K.}~\bibnamefont {Harii}}, \bibinfo
  {author} {\bibfnamefont {J.}~\bibnamefont {Ieda}}, \bibinfo {author}
  {\bibfnamefont {W.}~\bibnamefont {Koshibae}}, \bibinfo {author}
  {\bibfnamefont {K.}~\bibnamefont {Ando}}, \bibinfo {author} {\bibfnamefont
  {S.}~\bibnamefont {Maekawa}}, \ and\ \bibinfo {author} {\bibfnamefont
  {E.}~\bibnamefont {Saitoh}},\ }\href@noop {} {\bibfield  {journal} {\bibinfo
  {journal} {Nature}\ }\textbf {\bibinfo {volume} {455}},\ \bibinfo {pages}
  {778} (\bibinfo {year} {2008})}\BibitemShut {NoStop}%
\bibitem [{\citenamefont {Uchida}\ \emph {et~al.}(2010)\citenamefont {Uchida},
  \citenamefont {Adachi}, \citenamefont {Ota}, \citenamefont {Nakayama},
  \citenamefont {Maekawa},\ and\ \citenamefont {Saitoh}}]{Uchida:2010}%
  \BibitemOpen
  \bibfield  {author} {\bibinfo {author} {\bibfnamefont {K.}~\bibnamefont
  {Uchida}}, \bibinfo {author} {\bibfnamefont {H.}~\bibnamefont {Adachi}},
  \bibinfo {author} {\bibfnamefont {T.}~\bibnamefont {Ota}}, \bibinfo {author}
  {\bibfnamefont {H.}~\bibnamefont {Nakayama}}, \bibinfo {author}
  {\bibfnamefont {S.}~\bibnamefont {Maekawa}}, \ and\ \bibinfo {author}
  {\bibfnamefont {E.}~\bibnamefont {Saitoh}},\ }\href
  {http://scitation.aip.org/content/aip/journal/apl/97/17/10.1063/1.3507386}
  {\bibfield  {journal} {\bibinfo  {journal} {Appl. Phys. Lett.}\ }\textbf
  {\bibinfo {volume} {97}},\ \bibinfo {eid} {172505} (\bibinfo {year}
  {2010})}\BibitemShut {NoStop}%
\bibitem [{\citenamefont {Uchida}\ \emph {et~al.}(2014)\citenamefont {Uchida},
  \citenamefont {Ishida}, \citenamefont {Kikkawa}, \citenamefont {Kirihara},
  \citenamefont {Murakami},\ and\ \citenamefont {Saitoh}}]{Uchida:2014}%
  \BibitemOpen
  \bibfield  {author} {\bibinfo {author} {\bibfnamefont {K.}~\bibnamefont
  {Uchida}}, \bibinfo {author} {\bibfnamefont {M.}~\bibnamefont {Ishida}},
  \bibinfo {author} {\bibfnamefont {T.}~\bibnamefont {Kikkawa}}, \bibinfo
  {author} {\bibfnamefont {A.}~\bibnamefont {Kirihara}}, \bibinfo {author}
  {\bibfnamefont {T.}~\bibnamefont {Murakami}}, \ and\ \bibinfo {author}
  {\bibfnamefont {E.}~\bibnamefont {Saitoh}},\ }\href
  {http://stacks.iop.org/0953-8984/26/i=34/a=343202} {\bibfield  {journal}
  {\bibinfo  {journal} {J. Phys.: Condens. Matter}\ }\textbf {\bibinfo {volume}
  {26}},\ \bibinfo {pages} {343202} (\bibinfo {year} {2014})}\BibitemShut
  {NoStop}%
\bibitem [{\citenamefont {Huang}\ \emph {et~al.}(2012)\citenamefont {Huang},
  \citenamefont {Fan}, \citenamefont {Qu}, \citenamefont {Chen}, \citenamefont
  {Wang}, \citenamefont {Wu}, \citenamefont {Chen}, \citenamefont {Xiao},\ and\
  \citenamefont {Chien}}]{Huang:2012}%
  \BibitemOpen
  \bibfield  {author} {\bibinfo {author} {\bibfnamefont {S.~Y.}\ \bibnamefont
  {Huang}}, \bibinfo {author} {\bibfnamefont {X.}~\bibnamefont {Fan}}, \bibinfo
  {author} {\bibfnamefont {D.}~\bibnamefont {Qu}}, \bibinfo {author}
  {\bibfnamefont {Y.~P.}\ \bibnamefont {Chen}}, \bibinfo {author}
  {\bibfnamefont {W.~G.}\ \bibnamefont {Wang}}, \bibinfo {author}
  {\bibfnamefont {J.}~\bibnamefont {Wu}}, \bibinfo {author} {\bibfnamefont
  {T.~Y.}\ \bibnamefont {Chen}}, \bibinfo {author} {\bibfnamefont {J.~Q.}\
  \bibnamefont {Xiao}}, \ and\ \bibinfo {author} {\bibfnamefont {C.~L.}\
  \bibnamefont {Chien}},\ }\href@noop {} {\bibfield  {journal} {\bibinfo
  {journal} {Phys. Rev. Lett.}\ }\textbf {\bibinfo {volume} {109}},\ \bibinfo
  {pages} {107204} (\bibinfo {year} {2012})}\BibitemShut {NoStop}%
\bibitem [{\citenamefont {Bougiatioti}\ \emph {et~al.}(2017)\citenamefont
  {Bougiatioti}, \citenamefont {Klewe}, \citenamefont {Meier}, \citenamefont
  {Manos}, \citenamefont {Kuschel}, \citenamefont {Wollschl\"ager},
  \citenamefont {Bouchenoire}, \citenamefont {Brown}, \citenamefont
  {Schmalhorst}, \citenamefont {Reiss},\ and\ \citenamefont
  {Kuschel}}]{Bougiatioti:2017}%
  \BibitemOpen
  \bibfield  {author} {\bibinfo {author} {\bibfnamefont {P.}~\bibnamefont
  {Bougiatioti}}, \bibinfo {author} {\bibfnamefont {C.}~\bibnamefont {Klewe}},
  \bibinfo {author} {\bibfnamefont {D.}~\bibnamefont {Meier}}, \bibinfo
  {author} {\bibfnamefont {O.}~\bibnamefont {Manos}}, \bibinfo {author}
  {\bibfnamefont {O.}~\bibnamefont {Kuschel}}, \bibinfo {author} {\bibfnamefont
  {J.}~\bibnamefont {Wollschl\"ager}}, \bibinfo {author} {\bibfnamefont
  {L.}~\bibnamefont {Bouchenoire}}, \bibinfo {author} {\bibfnamefont {S.~D.}\
  \bibnamefont {Brown}}, \bibinfo {author} {\bibfnamefont {J.-M.}\ \bibnamefont
  {Schmalhorst}}, \bibinfo {author} {\bibfnamefont {G.}~\bibnamefont {Reiss}},
  \ and\ \bibinfo {author} {\bibfnamefont {T.}~\bibnamefont {Kuschel}},\
  }\href@noop {} {\bibfield  {journal} {\bibinfo  {journal} {arXiv:1702.05384}\
  } (\bibinfo {year} {2017})}\BibitemShut {NoStop}%
\bibitem [{\citenamefont {Meier}\ \emph {et~al.}(2013)\citenamefont {Meier},
  \citenamefont {Kuschel}, \citenamefont {Shen}, \citenamefont {Gupta},
  \citenamefont {Kikkawa}, \citenamefont {Uchida}, \citenamefont {Saitoh},
  \citenamefont {Schmalhorst},\ and\ \citenamefont {Reiss}}]{Meier:2013}%
  \BibitemOpen
  \bibfield  {author} {\bibinfo {author} {\bibfnamefont {D.}~\bibnamefont
  {Meier}}, \bibinfo {author} {\bibfnamefont {T.}~\bibnamefont {Kuschel}},
  \bibinfo {author} {\bibfnamefont {L.}~\bibnamefont {Shen}}, \bibinfo {author}
  {\bibfnamefont {A.}~\bibnamefont {Gupta}}, \bibinfo {author} {\bibfnamefont
  {T.}~\bibnamefont {Kikkawa}}, \bibinfo {author} {\bibfnamefont
  {K.}~\bibnamefont {Uchida}}, \bibinfo {author} {\bibfnamefont
  {E.}~\bibnamefont {Saitoh}}, \bibinfo {author} {\bibfnamefont {J.-M.}\
  \bibnamefont {Schmalhorst}}, \ and\ \bibinfo {author} {\bibfnamefont
  {G.}~\bibnamefont {Reiss}},\ }\href {\doibase 10.1103/PhysRevB.87.054421}
  {\bibfield  {journal} {\bibinfo  {journal} {Phys. Rev. B}\ }\textbf {\bibinfo
  {volume} {87}},\ \bibinfo {pages} {054421} (\bibinfo {year}
  {2013})}\BibitemShut {NoStop}%
\bibitem [{\citenamefont {Ramos}\ \emph {et~al.}(2013)\citenamefont {Ramos},
  \citenamefont {Kikkawa}, \citenamefont {Uchida}, \citenamefont {Adachi},
  \citenamefont {Lucas}, \citenamefont {Aguirre}, \citenamefont {Algarabel},
  \citenamefont {Morellón}, \citenamefont {Maekawa}, \citenamefont {Saitoh},\
  and\ \citenamefont {Ibarra}}]{Ramos:2013}%
  \BibitemOpen
  \bibfield  {author} {\bibinfo {author} {\bibfnamefont {R.}~\bibnamefont
  {Ramos}}, \bibinfo {author} {\bibfnamefont {T.}~\bibnamefont {Kikkawa}},
  \bibinfo {author} {\bibfnamefont {K.}~\bibnamefont {Uchida}}, \bibinfo
  {author} {\bibfnamefont {H.}~\bibnamefont {Adachi}}, \bibinfo {author}
  {\bibfnamefont {I.}~\bibnamefont {Lucas}}, \bibinfo {author} {\bibfnamefont
  {M.~H.}\ \bibnamefont {Aguirre}}, \bibinfo {author} {\bibfnamefont
  {P.}~\bibnamefont {Algarabel}}, \bibinfo {author} {\bibfnamefont
  {L.}~\bibnamefont {Morellón}}, \bibinfo {author} {\bibfnamefont
  {S.}~\bibnamefont {Maekawa}}, \bibinfo {author} {\bibfnamefont
  {E.}~\bibnamefont {Saitoh}}, \ and\ \bibinfo {author} {\bibfnamefont {M.~R.}\
  \bibnamefont {Ibarra}},\ }\href
  {http://scitation.aip.org/content/aip/journal/apl/102/7/10.1063/1.4793486}
  {\bibfield  {journal} {\bibinfo  {journal} {Appl. Phys. Lett.}\ }\textbf
  {\bibinfo {volume} {102}},\ \bibinfo {eid} {072413} (\bibinfo {year}
  {2013})}\BibitemShut {NoStop}%
\bibitem [{\citenamefont {Nakayama}\ \emph {et~al.}(2013)\citenamefont
  {Nakayama}, \citenamefont {Althammer}, \citenamefont {Chen}, \citenamefont
  {Uchida}, \citenamefont {Kajiwara}, \citenamefont {Kikuchi}, \citenamefont
  {Ohtani}, \citenamefont {Gepr\"ags}, \citenamefont {Opel}, \citenamefont
  {Takahashi}, \citenamefont {Gross}, \citenamefont {Bauer}, \citenamefont
  {Goennenwein},\ and\ \citenamefont {Saitoh}}]{Nakayama:2013}%
  \BibitemOpen
  \bibfield  {author} {\bibinfo {author} {\bibfnamefont {H.}~\bibnamefont
  {Nakayama}}, \bibinfo {author} {\bibfnamefont {M.}~\bibnamefont {Althammer}},
  \bibinfo {author} {\bibfnamefont {Y.-T.}\ \bibnamefont {Chen}}, \bibinfo
  {author} {\bibfnamefont {K.}~\bibnamefont {Uchida}}, \bibinfo {author}
  {\bibfnamefont {Y.}~\bibnamefont {Kajiwara}}, \bibinfo {author}
  {\bibfnamefont {D.}~\bibnamefont {Kikuchi}}, \bibinfo {author} {\bibfnamefont
  {T.}~\bibnamefont {Ohtani}}, \bibinfo {author} {\bibfnamefont
  {S.}~\bibnamefont {Gepr\"ags}}, \bibinfo {author} {\bibfnamefont
  {M.}~\bibnamefont {Opel}}, \bibinfo {author} {\bibfnamefont {S.}~\bibnamefont
  {Takahashi}}, \bibinfo {author} {\bibfnamefont {R.}~\bibnamefont {Gross}},
  \bibinfo {author} {\bibfnamefont {G.~E.~W.}\ \bibnamefont {Bauer}}, \bibinfo
  {author} {\bibfnamefont {S.~T.~B.}\ \bibnamefont {Goennenwein}}, \ and\
  \bibinfo {author} {\bibfnamefont {E.}~\bibnamefont {Saitoh}},\ }\href
  {\doibase 10.1103/PhysRevLett.110.206601} {\bibfield  {journal} {\bibinfo
  {journal} {Phys. Rev. Lett.}\ }\textbf {\bibinfo {volume} {110}},\ \bibinfo
  {pages} {206601} (\bibinfo {year} {2013})}\BibitemShut {NoStop}%
\bibitem [{\citenamefont {Chen}\ \emph {et~al.}(2013)\citenamefont {Chen},
  \citenamefont {Takahashi}, \citenamefont {Nakayama}, \citenamefont
  {Althammer}, \citenamefont {Goennenwein}, \citenamefont {Saitoh},\ and\
  \citenamefont {Bauer}}]{Chen:2013}%
  \BibitemOpen
  \bibfield  {author} {\bibinfo {author} {\bibfnamefont {Y.-T.}\ \bibnamefont
  {Chen}}, \bibinfo {author} {\bibfnamefont {S.}~\bibnamefont {Takahashi}},
  \bibinfo {author} {\bibfnamefont {H.}~\bibnamefont {Nakayama}}, \bibinfo
  {author} {\bibfnamefont {M.}~\bibnamefont {Althammer}}, \bibinfo {author}
  {\bibfnamefont {S.~T.~B.}\ \bibnamefont {Goennenwein}}, \bibinfo {author}
  {\bibfnamefont {E.}~\bibnamefont {Saitoh}}, \ and\ \bibinfo {author}
  {\bibfnamefont {G.~E.~W.}\ \bibnamefont {Bauer}},\ }\href {\doibase
  10.1103/PhysRevB.87.144411} {\bibfield  {journal} {\bibinfo  {journal} {Phys.
  Rev. B}\ }\textbf {\bibinfo {volume} {87}},\ \bibinfo {pages} {144411}
  (\bibinfo {year} {2013})}\BibitemShut {NoStop}%
\bibitem [{\citenamefont {Althammer}\ \emph {et~al.}(2013)\citenamefont
  {Althammer}, \citenamefont {Meyer}, \citenamefont {Nakayama}, \citenamefont
  {Schreier}, \citenamefont {Altmannshofer}, \citenamefont {Weiler},
  \citenamefont {Huebl}, \citenamefont {Gepr\"ags}, \citenamefont {Opel},
  \citenamefont {Gross}, \citenamefont {Meir}, \citenamefont {Klewe},
  \citenamefont {Kuschel}, \citenamefont {Schmalhorst}, \citenamefont {Reiss},
  \citenamefont {Shen}, \citenamefont {Gupta}, \citenamefont {Chen},
  \citenamefont {Bauer}, \citenamefont {Saitoh},\ and\ \citenamefont
  {Goennenwein}}]{Althammer:2013}%
  \BibitemOpen
  \bibfield  {author} {\bibinfo {author} {\bibfnamefont {M.}~\bibnamefont
  {Althammer}}, \bibinfo {author} {\bibfnamefont {S.}~\bibnamefont {Meyer}},
  \bibinfo {author} {\bibfnamefont {H.}~\bibnamefont {Nakayama}}, \bibinfo
  {author} {\bibfnamefont {M.}~\bibnamefont {Schreier}}, \bibinfo {author}
  {\bibfnamefont {S.}~\bibnamefont {Altmannshofer}}, \bibinfo {author}
  {\bibfnamefont {M.}~\bibnamefont {Weiler}}, \bibinfo {author} {\bibfnamefont
  {H.}~\bibnamefont {Huebl}}, \bibinfo {author} {\bibfnamefont
  {S.}~\bibnamefont {Gepr\"ags}}, \bibinfo {author} {\bibfnamefont
  {M.}~\bibnamefont {Opel}}, \bibinfo {author} {\bibfnamefont {R.}~\bibnamefont
  {Gross}}, \bibinfo {author} {\bibfnamefont {D.}~\bibnamefont {Meir}},
  \bibinfo {author} {\bibfnamefont {C.}~\bibnamefont {Klewe}}, \bibinfo
  {author} {\bibfnamefont {T.}~\bibnamefont {Kuschel}}, \bibinfo {author}
  {\bibfnamefont {J.-M.}\ \bibnamefont {Schmalhorst}}, \bibinfo {author}
  {\bibfnamefont {G.}~\bibnamefont {Reiss}}, \bibinfo {author} {\bibfnamefont
  {L.}~\bibnamefont {Shen}}, \bibinfo {author} {\bibfnamefont {A.}~\bibnamefont
  {Gupta}}, \bibinfo {author} {\bibfnamefont {Y.-T.}\ \bibnamefont {Chen}},
  \bibinfo {author} {\bibfnamefont {G.~E.~W.}\ \bibnamefont {Bauer}}, \bibinfo
  {author} {\bibfnamefont {E.}~\bibnamefont {Saitoh}}, \ and\ \bibinfo {author}
  {\bibfnamefont {S.~T.~B.}\ \bibnamefont {Goennenwein}},\ }\href@noop {}
  {\bibfield  {journal} {\bibinfo  {journal} {Phys. Rev. B}\ }\textbf {\bibinfo
  {volume} {87}},\ \bibinfo {pages} {224401} (\bibinfo {year}
  {2013})}\BibitemShut {NoStop}%
\bibitem [{\citenamefont {Pauthenet}(1952)}]{Pauthenet:1952}%
  \BibitemOpen
  \bibfield  {author} {\bibinfo {author} {\bibfnamefont {R.}~\bibnamefont
  {Pauthenet}},\ }\href@noop {} {\bibfield  {journal} {\bibinfo  {journal}
  {Ann. Phys.}\ }\textbf {\bibinfo {volume} {7}},\ \bibinfo {pages} {710}
  (\bibinfo {year} {1952})}\BibitemShut {NoStop}%
\bibitem [{\citenamefont {Dionne}(1988)}]{Dionne:1988}%
  \BibitemOpen
  \bibfield  {author} {\bibinfo {author} {\bibfnamefont {G.~F.}\ \bibnamefont
  {Dionne}},\ }\href@noop {} {\bibfield  {journal} {\bibinfo  {journal} {J.
  Appl. Phys.}\ }\textbf {\bibinfo {volume} {63}},\ \bibinfo {pages} {3777}
  (\bibinfo {year} {1988})}\BibitemShut {NoStop}%
\bibitem [{\citenamefont {Datta}\ \emph {et~al.}(2010)\citenamefont {Datta},
  \citenamefont {Kanuri}, \citenamefont {Karthik}, \citenamefont {Mazumdar},
  \citenamefont {Ma},\ and\ \citenamefont {Gupta}}]{Datta:2010}%
  \BibitemOpen
  \bibfield  {author} {\bibinfo {author} {\bibfnamefont {R.}~\bibnamefont
  {Datta}}, \bibinfo {author} {\bibfnamefont {S.}~\bibnamefont {Kanuri}},
  \bibinfo {author} {\bibfnamefont {V.}~\bibnamefont {Karthik}}, \bibinfo
  {author} {\bibfnamefont {D.}~\bibnamefont {Mazumdar}}, \bibinfo {author}
  {\bibfnamefont {J.~X.}\ \bibnamefont {Ma}}, \ and\ \bibinfo {author}
  {\bibfnamefont {A.}~\bibnamefont {Gupta}},\ }\href@noop {} {\bibfield
  {journal} {\bibinfo  {journal} {Appl. Phys. Lett.}\ }\textbf {\bibinfo
  {volume} {97}},\ \bibinfo {pages} {071907} (\bibinfo {year}
  {2010})}\BibitemShut {NoStop}%
\bibitem [{\citenamefont {Caltun}(2004)}]{Caltun:2004}%
  \BibitemOpen
  \bibfield  {author} {\bibinfo {author} {\bibfnamefont {O.~F.}\ \bibnamefont
  {Caltun}},\ }\href@noop {} {\bibfield  {journal} {\bibinfo  {journal} {J.
  Optoelectron. Adv. Mater.}\ }\textbf {\bibinfo {volume} {6}},\ \bibinfo
  {pages} {935} (\bibinfo {year} {2004})}\BibitemShut {NoStop}%
\bibitem [{\citenamefont {Hoppe}\ \emph {et~al.}(2015)\citenamefont {Hoppe},
  \citenamefont {D\"oring}, \citenamefont {Gorgoi}, \citenamefont {Cramm},\
  and\ \citenamefont {M\"uller}}]{Hoppe:2015}%
  \BibitemOpen
  \bibfield  {author} {\bibinfo {author} {\bibfnamefont {M.}~\bibnamefont
  {Hoppe}}, \bibinfo {author} {\bibfnamefont {S.}~\bibnamefont {D\"oring}},
  \bibinfo {author} {\bibfnamefont {M.}~\bibnamefont {Gorgoi}}, \bibinfo
  {author} {\bibfnamefont {S.}~\bibnamefont {Cramm}}, \ and\ \bibinfo {author}
  {\bibfnamefont {M.}~\bibnamefont {M\"uller}},\ }\href {\doibase
  10.1103/PhysRevB.91.054418} {\bibfield  {journal} {\bibinfo  {journal} {Phys.
  Rev. B}\ }\textbf {\bibinfo {volume} {91}},\ \bibinfo {pages} {054418}
  (\bibinfo {year} {2015})}\BibitemShut {NoStop}%
\bibitem [{\citenamefont {Li}\ \emph {et~al.}(2011)\citenamefont {Li},
  \citenamefont {Wang}, \citenamefont {Iliev}, \citenamefont {Klein},\ and\
  \citenamefont {Gupta}}]{Li:2011}%
  \BibitemOpen
  \bibfield  {author} {\bibinfo {author} {\bibfnamefont {N.}~\bibnamefont
  {Li}}, \bibinfo {author} {\bibfnamefont {Y.-H.~A.}\ \bibnamefont {Wang}},
  \bibinfo {author} {\bibfnamefont {M.~N.}\ \bibnamefont {Iliev}}, \bibinfo
  {author} {\bibfnamefont {T.~M.}\ \bibnamefont {Klein}}, \ and\ \bibinfo
  {author} {\bibfnamefont {A.}~\bibnamefont {Gupta}},\ }\href@noop {}
  {\bibfield  {journal} {\bibinfo  {journal} {Chem. Vap. Deposition}\ }\textbf
  {\bibinfo {volume} {17}},\ \bibinfo {pages} {261} (\bibinfo {year}
  {2011})}\BibitemShut {NoStop}%
\bibitem [{\citenamefont {Kuschel}\ \emph
  {et~al.}(2016{\natexlab{a}})\citenamefont {Kuschel}, \citenamefont {Bu\ss{}},
  \citenamefont {Spiess}, \citenamefont {Schemme}, \citenamefont
  {W\"ollermann}, \citenamefont {Balinski}, \citenamefont {N'Diaye},
  \citenamefont {Kuschel}, \citenamefont {Wollschl\"ager},\ and\ \citenamefont
  {Kuepper}}]{KuschelO:2016}%
  \BibitemOpen
  \bibfield  {author} {\bibinfo {author} {\bibfnamefont {O.}~\bibnamefont
  {Kuschel}}, \bibinfo {author} {\bibfnamefont {R.}~\bibnamefont {Bu\ss{}}},
  \bibinfo {author} {\bibfnamefont {W.}~\bibnamefont {Spiess}}, \bibinfo
  {author} {\bibfnamefont {T.}~\bibnamefont {Schemme}}, \bibinfo {author}
  {\bibfnamefont {J.}~\bibnamefont {W\"ollermann}}, \bibinfo {author}
  {\bibfnamefont {K.}~\bibnamefont {Balinski}}, \bibinfo {author}
  {\bibfnamefont {A.~T.}\ \bibnamefont {N'Diaye}}, \bibinfo {author}
  {\bibfnamefont {T.}~\bibnamefont {Kuschel}}, \bibinfo {author} {\bibfnamefont
  {J.}~\bibnamefont {Wollschl\"ager}}, \ and\ \bibinfo {author} {\bibfnamefont
  {K.}~\bibnamefont {Kuepper}},\ }\href {\doibase 10.1103/PhysRevB.94.094423}
  {\bibfield  {journal} {\bibinfo  {journal} {Phys. Rev. B}\ }\textbf {\bibinfo
  {volume} {94}},\ \bibinfo {pages} {094423} (\bibinfo {year}
  {2016}{\natexlab{a}})}\BibitemShut {NoStop}%
\bibitem [{\citenamefont {Klewe}\ \emph {et~al.}(2014)\citenamefont {Klewe},
  \citenamefont {Meinert}, \citenamefont {Boehnke}, \citenamefont {Kuepper},
  \citenamefont {Arenholz}, \citenamefont {Gupta}, \citenamefont {Schmalhorst},
  \citenamefont {Kuschel},\ and\ \citenamefont {Reiss}}]{Klewe:2014}%
  \BibitemOpen
  \bibfield  {author} {\bibinfo {author} {\bibfnamefont {C.}~\bibnamefont
  {Klewe}}, \bibinfo {author} {\bibfnamefont {M.}~\bibnamefont {Meinert}},
  \bibinfo {author} {\bibfnamefont {A.}~\bibnamefont {Boehnke}}, \bibinfo
  {author} {\bibfnamefont {K.}~\bibnamefont {Kuepper}}, \bibinfo {author}
  {\bibfnamefont {E.}~\bibnamefont {Arenholz}}, \bibinfo {author}
  {\bibfnamefont {A.}~\bibnamefont {Gupta}}, \bibinfo {author} {\bibfnamefont
  {J.-M.}\ \bibnamefont {Schmalhorst}}, \bibinfo {author} {\bibfnamefont
  {T.}~\bibnamefont {Kuschel}}, \ and\ \bibinfo {author} {\bibfnamefont
  {G.}~\bibnamefont {Reiss}},\ }\href
  {http://scitation.aip.org/content/aip/journal/jap/115/12/10.1063/1.4869400}
  {\bibfield  {journal} {\bibinfo  {journal} {J. Appl. Phys.}\ }\textbf
  {\bibinfo {volume} {115}},\ \bibinfo {eid} {123903} (\bibinfo {year}
  {2014})}\BibitemShut {NoStop}%
\bibitem [{\citenamefont {Kuschel}\ \emph
  {et~al.}(2016{\natexlab{b}})\citenamefont {Kuschel}, \citenamefont {Klewe},
  \citenamefont {Bougiatioti}, \citenamefont {Kuschel}, \citenamefont
  {Wollschläger}, \citenamefont {Bouchenoire}, \citenamefont {Brown},
  \citenamefont {Schmalhorst}, \citenamefont {Meier},\ and\ \citenamefont
  {Reiss}}]{Kuschel:2016}%
  \BibitemOpen
  \bibfield  {author} {\bibinfo {author} {\bibfnamefont {T.}~\bibnamefont
  {Kuschel}}, \bibinfo {author} {\bibfnamefont {C.}~\bibnamefont {Klewe}},
  \bibinfo {author} {\bibfnamefont {P.}~\bibnamefont {Bougiatioti}}, \bibinfo
  {author} {\bibfnamefont {O.}~\bibnamefont {Kuschel}}, \bibinfo {author}
  {\bibfnamefont {J.}~\bibnamefont {Wollschläger}}, \bibinfo {author}
  {\bibfnamefont {L.}~\bibnamefont {Bouchenoire}}, \bibinfo {author}
  {\bibfnamefont {S.~D.}\ \bibnamefont {Brown}}, \bibinfo {author}
  {\bibfnamefont {J.~M.}\ \bibnamefont {Schmalhorst}}, \bibinfo {author}
  {\bibfnamefont {D.}~\bibnamefont {Meier}}, \ and\ \bibinfo {author}
  {\bibfnamefont {G.}~\bibnamefont {Reiss}},\ }\href {\doibase
  10.1109/TMAG.2015.2512040} {\bibfield  {journal} {\bibinfo  {journal} {IEEE
  Trans. Magn.}\ }\textbf {\bibinfo {volume} {52}},\ \bibinfo {pages} {4500104}
  (\bibinfo {year} {2016}{\natexlab{b}})}\BibitemShut {NoStop}%
\bibitem [{\citenamefont {Kuschel}\ \emph {et~al.}(2015)\citenamefont
  {Kuschel}, \citenamefont {Klewe}, \citenamefont {Schmalhorst}, \citenamefont
  {Bertram}, \citenamefont {Kuschel}, \citenamefont {Schemme}, \citenamefont
  {Wollschl\"ager}, \citenamefont {Francoual}, \citenamefont {Strempfer},
  \citenamefont {Gupta}, \citenamefont {Meinert}, \citenamefont {G\"otz},
  \citenamefont {Meier},\ and\ \citenamefont {Reiss}}]{Kuschel:2015}%
  \BibitemOpen
  \bibfield  {author} {\bibinfo {author} {\bibfnamefont {T.}~\bibnamefont
  {Kuschel}}, \bibinfo {author} {\bibfnamefont {C.}~\bibnamefont {Klewe}},
  \bibinfo {author} {\bibfnamefont {J.-M.}\ \bibnamefont {Schmalhorst}},
  \bibinfo {author} {\bibfnamefont {F.}~\bibnamefont {Bertram}}, \bibinfo
  {author} {\bibfnamefont {O.}~\bibnamefont {Kuschel}}, \bibinfo {author}
  {\bibfnamefont {T.}~\bibnamefont {Schemme}}, \bibinfo {author} {\bibfnamefont
  {J.}~\bibnamefont {Wollschl\"ager}}, \bibinfo {author} {\bibfnamefont
  {S.}~\bibnamefont {Francoual}}, \bibinfo {author} {\bibfnamefont
  {J.}~\bibnamefont {Strempfer}}, \bibinfo {author} {\bibfnamefont
  {A.}~\bibnamefont {Gupta}}, \bibinfo {author} {\bibfnamefont
  {M.}~\bibnamefont {Meinert}}, \bibinfo {author} {\bibfnamefont
  {G.}~\bibnamefont {G\"otz}}, \bibinfo {author} {\bibfnamefont
  {D.}~\bibnamefont {Meier}}, \ and\ \bibinfo {author} {\bibfnamefont
  {G.}~\bibnamefont {Reiss}},\ }\href {\doibase 10.1103/PhysRevLett.115.097401}
  {\bibfield  {journal} {\bibinfo  {journal} {Phys. Rev. Lett.}\ }\textbf
  {\bibinfo {volume} {115}},\ \bibinfo {pages} {097401} (\bibinfo {year}
  {2015})}\BibitemShut {NoStop}%
\bibitem [{\citenamefont {Meier}\ \emph {et~al.}(2016)\citenamefont {Meier},
  \citenamefont {Kuschel}, \citenamefont {Meyer}, \citenamefont {Goennenwein},
  \citenamefont {Shen}, \citenamefont {Gupta}, \citenamefont {Schmalhorst},\
  and\ \citenamefont {Reiss}}]{MeierAIP:2016}%
  \BibitemOpen
  \bibfield  {author} {\bibinfo {author} {\bibfnamefont {D.}~\bibnamefont
  {Meier}}, \bibinfo {author} {\bibfnamefont {T.}~\bibnamefont {Kuschel}},
  \bibinfo {author} {\bibfnamefont {S.}~\bibnamefont {Meyer}}, \bibinfo
  {author} {\bibfnamefont {S.~T.~B.}\ \bibnamefont {Goennenwein}}, \bibinfo
  {author} {\bibfnamefont {L.}~\bibnamefont {Shen}}, \bibinfo {author}
  {\bibfnamefont {A.}~\bibnamefont {Gupta}}, \bibinfo {author} {\bibfnamefont
  {J.-M.}\ \bibnamefont {Schmalhorst}}, \ and\ \bibinfo {author} {\bibfnamefont
  {G.}~\bibnamefont {Reiss}},\ }\href@noop {} {\bibfield  {journal} {\bibinfo
  {journal} {AIP Adv.}\ }\textbf {\bibinfo {volume} {6}},\ \bibinfo {pages}
  {056302} (\bibinfo {year} {2016})}\BibitemShut {NoStop}%
\bibitem [{\citenamefont {Meier}\ \emph {et~al.}(2015)\citenamefont {Meier},
  \citenamefont {Reinhardt}, \citenamefont {van Straaten}, \citenamefont
  {Klewe}, \citenamefont {Althammer}, \citenamefont {Schreier}, \citenamefont
  {Goennenwein}, \citenamefont {Gupta}, \citenamefont {Schmid}, \citenamefont
  {Back}, \citenamefont {Schmalhorst}, \citenamefont {Kuschel},\ and\
  \citenamefont {Reiss}}]{Meier:2015}%
  \BibitemOpen
  \bibfield  {author} {\bibinfo {author} {\bibfnamefont {D.}~\bibnamefont
  {Meier}}, \bibinfo {author} {\bibfnamefont {D.}~\bibnamefont {Reinhardt}},
  \bibinfo {author} {\bibfnamefont {M.}~\bibnamefont {van Straaten}}, \bibinfo
  {author} {\bibfnamefont {C.}~\bibnamefont {Klewe}}, \bibinfo {author}
  {\bibfnamefont {M.}~\bibnamefont {Althammer}}, \bibinfo {author}
  {\bibfnamefont {M.}~\bibnamefont {Schreier}}, \bibinfo {author}
  {\bibfnamefont {S.~T.~B.}\ \bibnamefont {Goennenwein}}, \bibinfo {author}
  {\bibfnamefont {A.}~\bibnamefont {Gupta}}, \bibinfo {author} {\bibfnamefont
  {M.}~\bibnamefont {Schmid}}, \bibinfo {author} {\bibfnamefont {C.~H.}\
  \bibnamefont {Back}}, \bibinfo {author} {\bibfnamefont {J.-M.}\ \bibnamefont
  {Schmalhorst}}, \bibinfo {author} {\bibfnamefont {T.}~\bibnamefont
  {Kuschel}}, \ and\ \bibinfo {author} {\bibfnamefont {G.}~\bibnamefont
  {Reiss}},\ }\href@noop {} {\bibfield  {journal} {\bibinfo  {journal} {Nat.
  Commun.}\ }\textbf {\bibinfo {volume} {6}},\ \bibinfo {pages} {8211}
  (\bibinfo {year} {2015})}\BibitemShut {NoStop}%
\bibitem [{\citenamefont {Shan}\ \emph {et~al.}(2017)\citenamefont {Shan},
  \citenamefont {Bougiatioti}, \citenamefont {Liang}, \citenamefont {Reiss},
  \citenamefont {Kuschel},\ and\ \citenamefont {van Wees}}]{Juan:2017}%
  \BibitemOpen
  \bibfield  {author} {\bibinfo {author} {\bibfnamefont {J.}~\bibnamefont
  {Shan}}, \bibinfo {author} {\bibfnamefont {P.}~\bibnamefont {Bougiatioti}},
  \bibinfo {author} {\bibfnamefont {L.}~\bibnamefont {Liang}}, \bibinfo
  {author} {\bibfnamefont {G.}~\bibnamefont {Reiss}}, \bibinfo {author}
  {\bibfnamefont {T.}~\bibnamefont {Kuschel}}, \ and\ \bibinfo {author}
  {\bibfnamefont {B.~J.}\ \bibnamefont {van Wees}},\ }\href@noop {} {\bibfield
  {journal} {\bibinfo  {journal} {Appl. Phys. Lett.}\ }\textbf {\bibinfo
  {volume} {110}},\ \bibinfo {pages} {132406} (\bibinfo {year}
  {2017})}\BibitemShut {NoStop}%
\bibitem [{\citenamefont {Li}\ \emph {et~al.}(1991)\citenamefont {Li},
  \citenamefont {Fisher}, \citenamefont {Liu},\ and\ \citenamefont
  {Nevitt}}]{Li:1991}%
  \BibitemOpen
  \bibfield  {author} {\bibinfo {author} {\bibfnamefont {Z.}~\bibnamefont
  {Li}}, \bibinfo {author} {\bibfnamefont {E.~S.}\ \bibnamefont {Fisher}},
  \bibinfo {author} {\bibfnamefont {J.~Z.}\ \bibnamefont {Liu}}, \ and\
  \bibinfo {author} {\bibfnamefont {M.~V.}\ \bibnamefont {Nevitt}},\ }\href
  {\doibase 10.1007/BF02387728} {\bibfield  {journal} {\bibinfo  {journal} {J.
  Mater. Sci.}\ }\textbf {\bibinfo {volume} {26}},\ \bibinfo {pages} {2621}
  (\bibinfo {year} {1991})}\BibitemShut {NoStop}%
\bibitem [{\citenamefont {Harrison}(2005)}]{Harrison:2005}%
  \BibitemOpen
  \bibfield  {author} {\bibinfo {author} {\bibfnamefont {P.}~\bibnamefont
  {Harrison}},\ }\href@noop {} {\bibfield  {journal} {\bibinfo  {journal}
  {\textit{Quantum Wells, Wires and Dots: Theoretical and Computational Physics
  of Semiconductor Nanostructures}, 2nd ed. Wiley-Interscience, Chichester}\ }
  (\bibinfo {year} {2005})}\BibitemShut {NoStop}%
\bibitem [{\citenamefont {Fritsch}\ and\ \citenamefont
  {Ederer}(2010)}]{Fritsch:2010}%
  \BibitemOpen
  \bibfield  {author} {\bibinfo {author} {\bibfnamefont {D.}~\bibnamefont
  {Fritsch}}\ and\ \bibinfo {author} {\bibfnamefont {C.}~\bibnamefont
  {Ederer}},\ }\href {\doibase 10.1103/PhysRevB.82.104117} {\bibfield
  {journal} {\bibinfo  {journal} {Phys. Rev. B}\ }\textbf {\bibinfo {volume}
  {82}},\ \bibinfo {pages} {104117} (\bibinfo {year} {2010})}\BibitemShut
  {NoStop}%
\bibitem [{\citenamefont {Ma}\ \emph {et~al.}(2010)\citenamefont {Ma},
  \citenamefont {Mazumdar}, \citenamefont {Kim}, \citenamefont {Sato},
  \citenamefont {Bao},\ and\ \citenamefont {Gupta}}]{Ma:2010}%
  \BibitemOpen
  \bibfield  {author} {\bibinfo {author} {\bibfnamefont {J.~X.}\ \bibnamefont
  {Ma}}, \bibinfo {author} {\bibfnamefont {D.}~\bibnamefont {Mazumdar}},
  \bibinfo {author} {\bibfnamefont {G.}~\bibnamefont {Kim}}, \bibinfo {author}
  {\bibfnamefont {H.}~\bibnamefont {Sato}}, \bibinfo {author} {\bibfnamefont
  {N.~Z.}\ \bibnamefont {Bao}}, \ and\ \bibinfo {author} {\bibfnamefont
  {A.}~\bibnamefont {Gupta}},\ }\href@noop {} {\bibfield  {journal} {\bibinfo
  {journal} {J. Appl. Phys.}\ }\textbf {\bibinfo {volume} {108}},\ \bibinfo
  {pages} {063917} (\bibinfo {year} {2010})}\BibitemShut {NoStop}%
\bibitem [{\citenamefont {Gantmakher}\ and\ \citenamefont
  {Lucia}(2005)}]{Gantmakher:2005}%
  \BibitemOpen
  \bibfield  {author} {\bibinfo {author} {\bibfnamefont {V.~F.}\ \bibnamefont
  {Gantmakher}}\ and\ \bibinfo {author} {\bibfnamefont {I.~M.}\ \bibnamefont
  {Lucia}},\ }\href@noop {} {\bibfield  {journal} {\bibinfo  {journal}
  {Electrons and disorder in solids, Oxford : Clarendon Press ; Oxford ; New
  York : Oxford University Press, Oxford}\ } (\bibinfo {year}
  {2005})}\BibitemShut {NoStop}%
\bibitem [{\citenamefont {Shklovskii}\ and\ \citenamefont
  {Efros}(1984)}]{ES:1984}%
  \BibitemOpen
  \bibfield  {author} {\bibinfo {author} {\bibfnamefont {B.~I.}\ \bibnamefont
  {Shklovskii}}\ and\ \bibinfo {author} {\bibfnamefont {A.~L.}\ \bibnamefont
  {Efros}},\ }\href@noop {} {\bibfield  {journal} {\bibinfo  {journal}
  {\textit{Electronic Properties of Doped Semiconductors}, Springer, Berlin}\ }
  (\bibinfo {year} {1984})}\BibitemShut {NoStop}%
\bibitem [{\citenamefont {Mott}(1969)}]{Mott:1969}%
  \BibitemOpen
  \bibfield  {author} {\bibinfo {author} {\bibfnamefont {N.~F.}\ \bibnamefont
  {Mott}},\ }\href@noop {} {\bibfield  {journal} {\bibinfo  {journal} {Contemp.
  Phys.}\ }\textbf {\bibinfo {volume} {10}},\ \bibinfo {pages} {125} (\bibinfo
  {year} {1969})}\BibitemShut {NoStop}%
\bibitem [{\citenamefont {Lord}\ and\ \citenamefont
  {Parker}(1960)}]{Lord:1960}%
  \BibitemOpen
  \bibfield  {author} {\bibinfo {author} {\bibfnamefont {H.}~\bibnamefont
  {Lord}}\ and\ \bibinfo {author} {\bibfnamefont {R.}~\bibnamefont {Parker}},\
  }\href@noop {} {\bibfield  {journal} {\bibinfo  {journal} {Nature}\ }\textbf
  {\bibinfo {volume} {188}},\ \bibinfo {pages} {929} (\bibinfo {year}
  {1960})}\BibitemShut {NoStop}%
\bibitem [{\citenamefont {Austin}\ and\ \citenamefont
  {Elwell}(1970)}]{Austin:1970}%
  \BibitemOpen
  \bibfield  {author} {\bibinfo {author} {\bibfnamefont {I.~G.}\ \bibnamefont
  {Austin}}\ and\ \bibinfo {author} {\bibfnamefont {D.}~\bibnamefont
  {Elwell}},\ }\href {\doibase 10.1080/00107517008202186} {\bibfield  {journal}
  {\bibinfo  {journal} {Contemp. Phys.}\ }\textbf {\bibinfo {volume} {11}},\
  \bibinfo {pages} {455} (\bibinfo {year} {1970})}\BibitemShut {NoStop}%
\bibitem [{\citenamefont {Ponpandian}\ \emph {et~al.}(2002)\citenamefont
  {Ponpandian}, \citenamefont {Balaya},\ and\ \citenamefont
  {Narayanasamy}}]{Ponpandian:2002}%
  \BibitemOpen
  \bibfield  {author} {\bibinfo {author} {\bibfnamefont {N.}~\bibnamefont
  {Ponpandian}}, \bibinfo {author} {\bibfnamefont {P.}~\bibnamefont {Balaya}},
  \ and\ \bibinfo {author} {\bibfnamefont {A.}~\bibnamefont {Narayanasamy}},\
  }\href {http://stacks.iop.org/0953-8984/14/i=12/a=311} {\bibfield  {journal}
  {\bibinfo  {journal} {J. Phys. Condens. Matter}\ }\textbf {\bibinfo {volume}
  {14}},\ \bibinfo {pages} {3221} (\bibinfo {year} {2002})}\BibitemShut
  {NoStop}%
\bibitem [{\citenamefont {Rolin}\ \emph {et~al.}(2017)\citenamefont {Rolin},
  \citenamefont {Kang}, \citenamefont {Lee}, \citenamefont {Borghs},
  \citenamefont {Heremans},\ and\ \citenamefont {Genoe}}]{Rolin:2017}%
  \BibitemOpen
  \bibfield  {author} {\bibinfo {author} {\bibfnamefont {C.}~\bibnamefont
  {Rolin}}, \bibinfo {author} {\bibfnamefont {E.}~\bibnamefont {Kang}},
  \bibinfo {author} {\bibfnamefont {J.-H.}\ \bibnamefont {Lee}}, \bibinfo
  {author} {\bibfnamefont {G.}~\bibnamefont {Borghs}}, \bibinfo {author}
  {\bibfnamefont {P.}~\bibnamefont {Heremans}}, \ and\ \bibinfo {author}
  {\bibfnamefont {J.}~\bibnamefont {Genoe}},\ }\href
  {http://dx.doi.org/10.1038/ncomms14975} {\bibfield  {journal} {\bibinfo
  {journal} {Nat. Commun.}\ }\textbf {\bibinfo {volume} {8}},\ \bibinfo {pages}
  {14975} (\bibinfo {year} {2017})}\BibitemShut {NoStop}%
\bibitem [{\citenamefont {Meinert}\ and\ \citenamefont
  {Reiss}(2014)}]{Meinert:2014}%
  \BibitemOpen
  \bibfield  {author} {\bibinfo {author} {\bibfnamefont {M.}~\bibnamefont
  {Meinert}}\ and\ \bibinfo {author} {\bibfnamefont {G.}~\bibnamefont
  {Reiss}},\ }\href {http://stacks.iop.org/0953-8984/26/i=11/a=115503}
  {\bibfield  {journal} {\bibinfo  {journal} {J. Phys.: Condens. Matter}\
  }\textbf {\bibinfo {volume} {26}},\ \bibinfo {pages} {115503} (\bibinfo
  {year} {2014})}\BibitemShut {NoStop}%
\bibitem [{\citenamefont {Holinsworth}\ \emph {et~al.}(2013)\citenamefont
  {Holinsworth}, \citenamefont {Mazumdar}, \citenamefont {Sims}, \citenamefont
  {Sun}, \citenamefont {Yurtisigi}, \citenamefont {Sarker}, \citenamefont
  {Gupta}, \citenamefont {Butler},\ and\ \citenamefont
  {Musfeldt}}]{Holinsworth:2013}%
  \BibitemOpen
  \bibfield  {author} {\bibinfo {author} {\bibfnamefont {B.~S.}\ \bibnamefont
  {Holinsworth}}, \bibinfo {author} {\bibfnamefont {D.}~\bibnamefont
  {Mazumdar}}, \bibinfo {author} {\bibfnamefont {H.}~\bibnamefont {Sims}},
  \bibinfo {author} {\bibfnamefont {Q.-C.}\ \bibnamefont {Sun}}, \bibinfo
  {author} {\bibfnamefont {M.~K.}\ \bibnamefont {Yurtisigi}}, \bibinfo {author}
  {\bibfnamefont {S.~K.}\ \bibnamefont {Sarker}}, \bibinfo {author}
  {\bibfnamefont {A.}~\bibnamefont {Gupta}}, \bibinfo {author} {\bibfnamefont
  {W.~H.}\ \bibnamefont {Butler}}, \ and\ \bibinfo {author} {\bibfnamefont
  {J.~L.}\ \bibnamefont {Musfeldt}},\ }\href@noop {} {\bibfield  {journal}
  {\bibinfo  {journal} {Appl. Phys. Lett.}\ }\textbf {\bibinfo {volume}
  {103}},\ \bibinfo {pages} {082406} (\bibinfo {year} {2013})}\BibitemShut
  {NoStop}%
\bibitem [{\citenamefont {Sun}\ \emph {et~al.}(2012)\citenamefont {Sun},
  \citenamefont {Sims}, \citenamefont {Mazumdar}, \citenamefont {Ma},
  \citenamefont {Holinsworth}, \citenamefont {O'Neal}, \citenamefont {Kim},
  \citenamefont {Butler}, \citenamefont {Gupta},\ and\ \citenamefont
  {Musfeldt}}]{Sun:2012}%
  \BibitemOpen
  \bibfield  {author} {\bibinfo {author} {\bibfnamefont {Q.~C.}\ \bibnamefont
  {Sun}}, \bibinfo {author} {\bibfnamefont {H.}~\bibnamefont {Sims}}, \bibinfo
  {author} {\bibfnamefont {D.}~\bibnamefont {Mazumdar}}, \bibinfo {author}
  {\bibfnamefont {J.~X.}\ \bibnamefont {Ma}}, \bibinfo {author} {\bibfnamefont
  {B.~S.}\ \bibnamefont {Holinsworth}}, \bibinfo {author} {\bibfnamefont
  {K.~R.}\ \bibnamefont {O'Neal}}, \bibinfo {author} {\bibfnamefont
  {G.}~\bibnamefont {Kim}}, \bibinfo {author} {\bibfnamefont {W.~H.}\
  \bibnamefont {Butler}}, \bibinfo {author} {\bibfnamefont {A.}~\bibnamefont
  {Gupta}}, \ and\ \bibinfo {author} {\bibfnamefont {J.~L.}\ \bibnamefont
  {Musfeldt}},\ }\href {\doibase 10.1103/PhysRevB.86.205106} {\bibfield
  {journal} {\bibinfo  {journal} {Phys. Rev. B}\ }\textbf {\bibinfo {volume}
  {86}},\ \bibinfo {pages} {205106} (\bibinfo {year} {2012})}\BibitemShut
  {NoStop}%
\end{thebibliography}%

\end{document}